\documentclass[aps,reprint,nofootinbib,preprintnumbers,twocolumn,floatfix,superscriptaddress,pra,showpacs]{revtex4-1}
\usepackage{physics}
\usepackage{natbib}
\usepackage{amsmath,amssymb}
\usepackage{graphicx}
\usepackage[dvipsnames]{xcolor}
\usepackage[caption=false]{subfig}
\usepackage{float}
\usepackage[pdftex]{hyperref} 
\usepackage{wasysym}
\usepackage{enumitem}
\usepackage{makecell}
\usepackage{array}

\begin{document}
	\title{Quantum Networks with Deterministic Spin-Photon Interfaces}
	\author{J. Borregaard}
	\affiliation{QMATH, Department of Mathematical Sciences, University of Copenhagen, 2100 Copenhagen \O, Denmark}
	\author{A. S. S\o rensen}
	\affiliation{Center for Hybrid Quantum Networks (Hy-Q), The Niels Bohr Institute, University of Copenhagen, Blegdamsvej 17, DK-2100 Copenhagen \O, Denmark}
	\author{P.  Lodahl}
	\affiliation{Center for Hybrid Quantum Networks (Hy-Q), The Niels Bohr Institute, University of Copenhagen, Blegdamsvej 17, DK-2100 Copenhagen \O, Denmark}
	\date{\today}
	
	\begin{abstract}
We consider how recent experimental progress on deterministic solid state spin-photon interfaces enable the construction of a number of key elements of quantum networks. After reviewing some of the recent experimental achievements, we discuss their integration into Bell state analyzers, quantum non-demolition detection, and photonic cluster state generation. Finally, we outline how these elements can be used for long-distance entanglement generation and quantum key distribution in a quantum network.
	\end{abstract}
	\maketitle
\section{Introduction}
Over the past decades, the counterintuitive and mind-boggling features of quantum mechanics have moved from the stage of theoretical "Gedanken Experiments" to being the effects underlying the development of a whole new range of quantum technologies. The promises of ultra-sensitive metrology~\cite{Giovannetti2011}, powerful quantum computers~\cite{Ladd2010}, and new cryptographic primitives~\cite{Gisin2007} have spurred substantial interest worldwide. A plethora of experimental platforms are currently being pursued as possible hardware candidates each with their different strengths and weaknesses. The hardware of choice strongly depends on  the application in mind. There have been impressive developments of trapped ions~\cite{Zhang2017}, Rydberg atoms~\cite{Bernien2017} and superconducting qubits~\cite{Wendin2017} for quantum computation, while quantum communication applications such as quantum key distribution (QKD) require optical photons that could be generated by single atoms~\cite{Ritter2012}, solid-state defects~\cite{Schroder2016}, or quantum dots~\cite{Lodahl2015}. This has spurred substantial experimental progress towards deterministic solid-state spin-photon interfaces~\cite{Lodahl2018}. In this progress report, we discuss some of these experimental developments and consider how they may enable the implementation of protocols for quantum key distribution and quantum networks in general.

\section{The deterministic spin-photon interface} \label{sec:hardware}

\begingroup
\begin{table*} [t]
\begin{tabular}{|p{3cm}|p{6 cm}|p{6 cm}|}
\hline
Notation & Parameter & Relevance \\ \hline
\begin{itemize}[noitemsep,nolistsep,leftmargin=*] \item[]$\gamma_{\text{rad}}$ \end{itemize}& \begin{itemize}[noitemsep,nolistsep,leftmargin=*] \item[] Free-space radiative decay rate of emitter. \end{itemize} & \begin{itemize}[noitemsep,nolistsep,leftmargin=*]
\item[-] Loss of photons.
\item[-] Efficiency of spin-photon interface.\end{itemize} \\ \hline
\begin{itemize}[noitemsep,nolistsep,leftmargin=*]\item[]$\gamma_{\text{nonrad}}$\end{itemize} &\begin{itemize}[noitemsep,nolistsep,leftmargin=*] \item[] Non-radiative decay rate of emitter. \end{itemize} & \begin{itemize}[noitemsep,nolistsep,leftmargin=*]
\item[-]Efficiency of spin-photon interface.\end{itemize}\\ \hline
\begin{itemize}[noitemsep,nolistsep,leftmargin=*] \item[]$\gamma_{\text{dp}}$ \end{itemize}& \begin{itemize}[noitemsep,nolistsep,leftmargin=*] \item[]Pure dephasing rate. \end{itemize} &  \begin{itemize}[noitemsep,nolistsep,leftmargin=*]
\item[-] Indistinguishability of photons.
\item[-] Fidelity of protocols.\end{itemize}\\ \hline
\begin{itemize}[noitemsep,nolistsep,leftmargin=*]\item[]$\Gamma$\end{itemize} &\begin{itemize}[noitemsep,nolistsep,leftmargin=*] \item[] Total decay rate of emitter. \end{itemize} & \begin{itemize}[noitemsep,nolistsep,leftmargin=*]
\item[-]Bandwidth of spin-photon interface.\end{itemize}\\ \hline
\begin{itemize}[noitemsep,nolistsep,leftmargin=*]\item[]$T_{\text{coh}}$ \end{itemize} & \begin{itemize}[noitemsep,nolistsep,leftmargin=*] \item[]Spin coherence time. \end{itemize} & \begin{itemize}[noitemsep,nolistsep,leftmargin=*]
\item[-] Quantum memory time.\end{itemize} \\ \hline
\begin{itemize}[noitemsep,nolistsep,leftmargin=*] \item[]$\Delta\omega$ \end{itemize}&\begin{itemize}[noitemsep,nolistsep,leftmargin=*]\item[] Inhomogeneous broadening. \end{itemize}&\begin{itemize}[noitemsep,nolistsep,leftmargin=*]
\item[-]  Stability of optical transition.
\item[-] Indistinguishability of photons.
\item[-] Fidelity of protocols.
\end{itemize} \\ \hline
\begin{itemize}[noitemsep,nolistsep,leftmargin=*]\item[]$\beta=\frac{\Gamma_{\text{1d}}}{\Gamma}$\end{itemize} & \begin{itemize}[noitemsep,nolistsep,leftmargin=*]\item[]Ratio between the optical decay rate into a waveguide ($\Gamma_{\text{1d}}$) and the total decay rate of emitter. \end{itemize}&  \begin{itemize}[noitemsep,nolistsep,leftmargin=*]
\item[-] Efficiency of spin-photon interface.
\end{itemize}   \\ \hline
\begin{itemize}[noitemsep,nolistsep,leftmargin=*]\item[]$\beta_{\text{coh}}=\frac{\Gamma_{\text{1d}}}{\Gamma+\gamma_{\text{dp}}}$ \end{itemize} & \begin{itemize}[noitemsep,nolistsep,leftmargin=*] \item[] Ratio between optical decay rate into a waveguide and the total decay rate including the pure dephasing rate of the emitter.  \end{itemize}&  \begin{itemize}[noitemsep,nolistsep,leftmargin=*]
\item[-] Efficiency of coherent photon generation.
\item[-] Fidelity of protocols.
\end{itemize}   \\ \hline
\begin{itemize}[noitemsep,nolistsep,leftmargin=*]\item[]$C=$\newline$\frac{4\abs{g}^2}{\kappa(\gamma_{\text{rad}}+\gamma_{\text{nonrad}})}$\end{itemize} & \begin{itemize}[noitemsep,nolistsep,leftmargin=*]\item[]Cooperativity of an emitter coupled to a cavity, where $g$ is the single photon Rabi frequency and $\kappa$ is the total decay rate of the cavity field. \end{itemize} & \begin{itemize}[noitemsep,nolistsep,leftmargin=*]
\item[-] Ratio of decay into the cavity to undesired decay.
\item[-] Efficiency of spin-photon interface.
\end{itemize} \\\hline
\begin{itemize}[noitemsep,nolistsep,leftmargin=*]\item[]$C_{\text{coh}}=$\newline$\tfrac{4\abs{g}^2}{\kappa(\gamma_{\text{rad}}+\gamma_{\text{nonrad}}+\gamma_{\text{dp}})}$\end{itemize} & \begin{itemize}[noitemsep,nolistsep,leftmargin=*]\item[]Cooperativity of an emitter coupled to a cavity including the pure dephasing rate of the emitter. \end{itemize} & \begin{itemize}[noitemsep,nolistsep,leftmargin=*]
\item[-] Ratio of coherent decay into the cavity to undesired decay.
\item[-] Efficiency of coherent photon generation.
\item[-] Fidelity of protocols.
\end{itemize} \\\hline
\begin{itemize}[noitemsep,nolistsep,leftmargin=*]\item[]$\eta_{\text{in}}/\eta_\text{out}$ \end{itemize}& \begin{itemize}[noitemsep,nolistsep,leftmargin=*]\item[]Input/output coupling efficiency of light. \end{itemize} & \begin{itemize}[noitemsep,nolistsep,leftmargin=*]
\item[-] Efficiency of quantum operations.
\item[-] Loss errors.
\end{itemize} \\ \hline
\end{tabular}
\caption{Parameters characterizing a spin-photon interface and their relevance for the quantum-information protocols discussed in the main text.  \label{tab:table1}}
\end{table*}
\endgroup

\begin{figure} [t]
\centering
\includegraphics[width=0.45\textwidth]{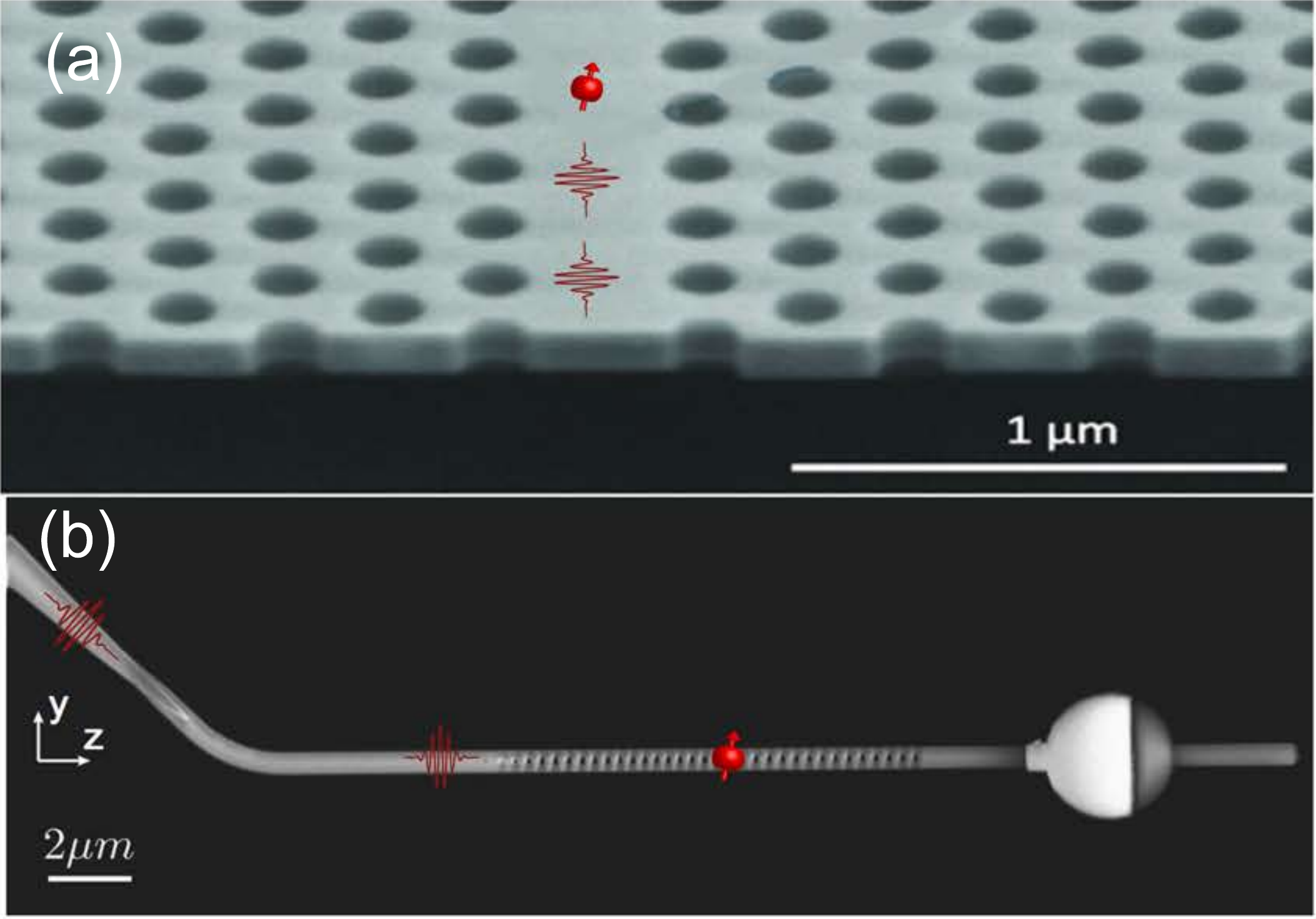}
\caption{Scanning-electron microscope image of photonic crystal (PC) structures with single spins and photons sketched on top. A PC waveguide is shown in (a), which can allow for near-unity emission of single photons from a spin system into a propagating guided mode. A PC cavity evanescently coupled to an optical fiber is shown in (b). The images are reproduced (adapted) with permission (a): Ref.~\cite{Lodahl2015}, 2015, American Physical Society and (b): Ref.~\cite{Tiecke2014}, 2014, Springer Nature.}
\label{fig:figurePC}
\end{figure}

Optical photons are the carriers of choice for distributing quantum information over long distances, since photons can propagate with low-loss through optical fibers and encounter negligible thermal noise even at room temperature. On the other hand, for application in quantum information processing it is essential to  be able to store and process the information encoded in the photons.  As a consequence, an efficient interface between light and matter is an essential building block. To this end, the atomic-physics community has pioneered cavity quantum electrodynamics as an approach for interfacing single atoms and single photons by strongly enhancing the electromagnetic field in a resonator~\cite{Kimble1998,Thompson2013,Reiserer2015}. More recently, solid-state implementations have been developed where single atoms are replaced by solid-state quantum emitters such as quantum dots~\cite{Warburton2013,Senellart2017} or vacancy centers in diamond~\cite{Doherty2013}. Furthermore, the development of nanophotonics implies that advanced devices can be fabricated where light-matter interaction is precisely tailored~\cite{Lodahl2015}. Consequently, it is today possible to engineer an almost deterministic interface between a single photon and a single quantum emitter, by creating conditions where the emitter is preferentially coupled to a single mode of a cavity or a waveguide (see Fig.~\ref{fig:figurePC}).  If the ground state of the quantum emitter consists of a coherent spin, the interface comprises a quantum memory enabling advanced quantum functionalities.
Various experimental implementations of spin-photon interfaces have been studied using, e.g., single trapped atoms in cavities~\cite{Birnbaum2005,Boozer2007,Tiecke2014,Hacker2016}, Silicon (SiV) or Nitrogen (NV) vacancy centers in diamond~\cite{Bernien2013,Sipahigil2016,Kalb2017} or self-assembled quantum dots in gallium arsenide~\cite{Arcari2014,Delteil2015,Kuhlmann2015,De-Santis2017,Senellart2017,Javadi2018}. Here our main focus will be on implementations where the photon-emitter coupling efficiency is near unity and highly coherent, which is the limit where spin and photon become deterministically coupled.

Efficient spin-photon interfaces can be implemented in either cavity or waveguide geometries, cf. Fig. \ref{fig:figurePC}, corresponding to the case where the emitter is coupled to a localized or a travelling photon. Similar functionalities can in general be implemented on both platforms. In open waveguide geometries, however, the ability to engineer a chiral light-matter coupling can lead to new opportunities for spin-path photon entanglement, integrated quantum photonic circuits, and multi-emitter coupling~\cite{Coles2016,Mahmoodian2016,Lodahl2017}. The relevant figures-of-merit characterizing the spin-photon interface are summarized in Table~\ref{tab:table1}. In essence, the radiative emitter decay time should be short in order to rapidly generate photons, the spin coherence time long to generate high-fidelity multi-photon entangled states, and any homogenous and inhomogeneous broadening should be reduced in order to obtain indistinguishable photons. For waveguide implementations, the $\beta$-factor is the essential parameter that characterizes the probability to generate a photon in the desired mode. For instance, in QKD applications, the relevant figure-of-merit is the probability to get a photon into an optical fiber $\beta\eta_{\text{out}}$, which is determined by the outcoupling efficiency  $\eta_{\text{out}}$ and the capture probability  of the nanostructure $\beta$ (see Tab.~\ref{tab:table1}). 

For applications requiring the interference of different photons or interaction of single photons with the quantum emitter, any decoherence processes become relevant. The pure dephasing rate $\gamma_{\text{dp}}$ signifies the broadening due to fast decoherence processes and is important for characterizing an emitter. It is therefore often convenient to introduce the coherent $\beta_{\text{coh}}$ factor, which includes the pure dephasing rate (see Tab.~\ref{tab:table1}). When considering photon interference between different quantum emitters, the inhomogeneous broadening of the emitters $\Delta \omega$ needs to be considered as well. 

For applications involving storage of photons, time reversal arguments can be used to show~\cite{Gorshkov2007,Chang2007} that the probability for an emitter to absorb an incoming photon is given by the same efficiency as the probability to emit a photon into the desired mode. Hence the beta-factor also plays an important role for storage. 

For several applications in quantum information processing it is also favorable to exploit the effective non-linear interaction between photons induced by the emitters.  Ultimately such non-linearity arises from the fact that two photons cannot be absorbed by the same emitter simultaneously and is thus determined by the $\beta$ factor. Note, however, that various decoherence processes may leak qubit information to the environment and should therefore also be carefully considered for such applications. 

The above discussion has been phrased in the language of wavequide interfaces where the $\beta$ factor is the most important quantity. For implementations based on optical cavities, the figure of merit for the quality of the interfaces is typically expressed in terms of the cooperativity $C=4\abs{g}^2/(\kappa(\gamma_{\text{rad}}+\gamma_{\text{nonrad}}))$, where $g$ is the single photon Rabi frequency of the cavity coupled transition and $\kappa$ is the total decay rate of the cavity field. The free space decay rate and non-radiative decay rate is denoted $\gamma_{\text{rad}}$ and $\gamma_{\text{nonrad}}$, respectively. For broadband cavities, the cooperativity expresses the ratio of the cavity induced decay rate to the decay rate in the absence of the cavity. Hence the equivalent of the $\beta$ factor, the probability to decay through the cavity field, can be expressed as $C/(1+C)$. As with the $\beta$ factor, it is also convenient to introduce a coherent cooperativity $C_{\text{coh}}$ that includes the pure dephasing rate of the emitter (see Tab.~\ref{tab:table1}). With this identification, the functionalities of interfaces based on waveguides and cavities becomes almost identical, with only minor differences between them. The protocols described below are thus applicable for both implementations, although there may be differences in the required linear optical elements surrounding the interface.  

\section{Theoretical building blocks} \label{sec:elements}
The access to deterministic spin-photon interfaces opens up new routes to realize some basic elements of a quantum network. In this section, we will discuss how optical Bell state analyzers, photonic quantum non-demolition detectors, and photonic cluster-state generation may be realized with such hardware. In Sec.~\ref{sec:repeaters}, we outline how these elements are crucial to a number of proposals for entanglement  and long-distance quantum key distribution in quantum networks.

\subsection{Optical Bell state analyzer} \label{sec:Bell}
A Bell state analyzer is a device that allows to measure two qubits in the Bell basis consisting of the four Bell states $\ket{\phi^{\pm}}=(\ket{00}\pm\ket{11})/\sqrt{2}$ and $\ket{\psi^{\pm}}=(\ket{01}\pm\ket{10})/\sqrt{2}$ for logical states $\ket{0}$ and $\ket{1}$. Such a device can be used for both heralded entanglement generation and entanglement swapping in quantum repeaters. In addition, a Bell state analyzer can also be used for fusion gates in cluster state generation, as we discuss below and for estimation of purity of a quantum state~\cite{Lutkenhaus1999}. 

A deterministic Bell analyzer is not possible with linear optics elements~\cite{Lutkenhaus1999} but probabilistic versions have been proposed and realized for photonic qubits ~\cite{Weinfurter1994,Braunstein1995,Michler1996}. Without auxiliary photons, the maximum success probability is $50\%$~\cite{Braunstein1995,Calsamiglia2001} but allowing for auxiliary multi-photon states can enable near-deterministic Bell analyzers based on linear optics~\cite{Grice2011,Ewert2014}. While a success probability of 75\% is possible using only 4 auxiliary single photons~\cite{Ewert2014}, this approach, in general, requires the generation of multi-photon entangled states. For instance, an entangled state of 30 photons is needed to reach a success probability of $\sim97\%$ using the scheme of Ref.~\cite{Grice2011}. 

\begin{figure} [t]
\centering
\includegraphics[width=0.45\textwidth]{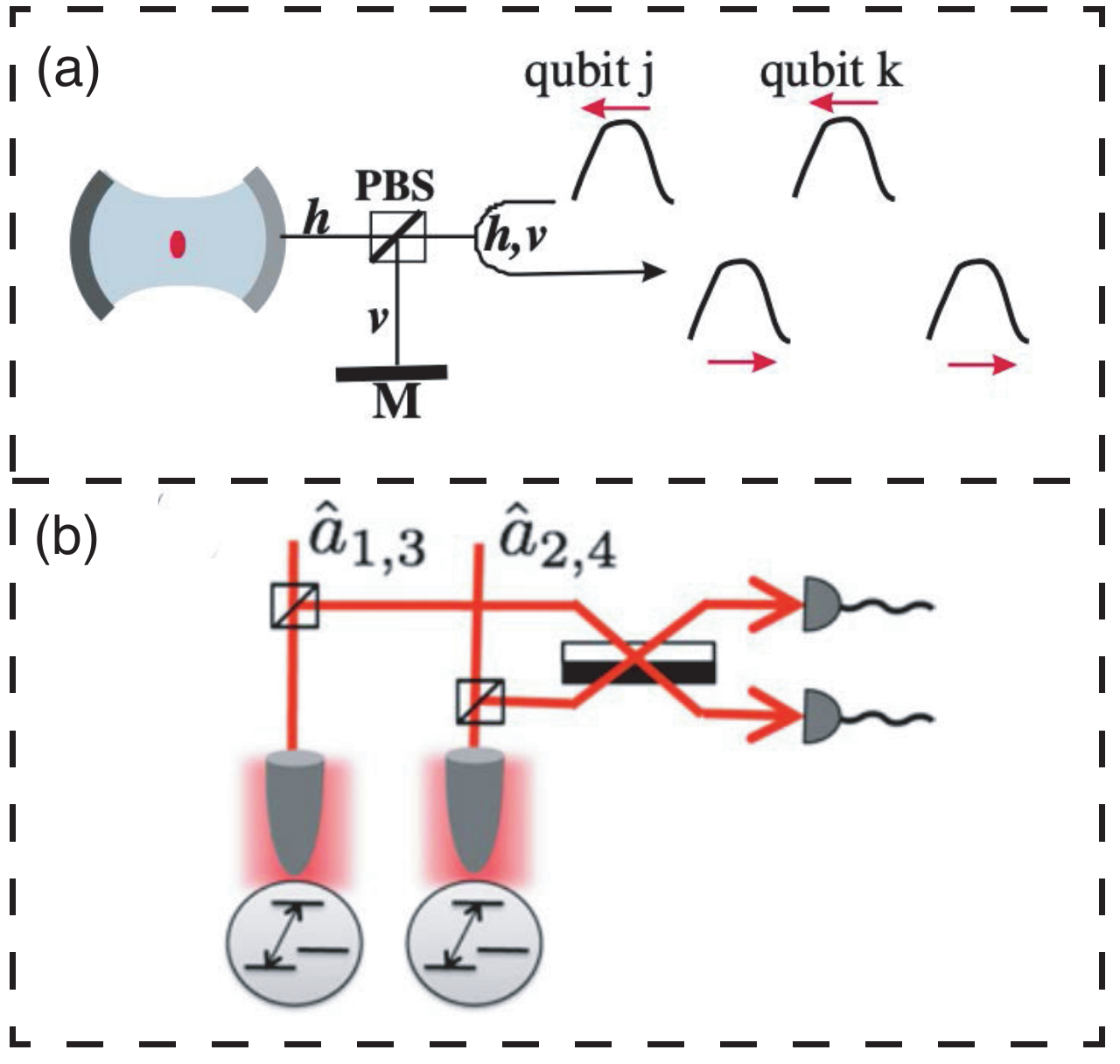}
\caption{(a) Schematic setup for the optical CZ-gate of Ref.~\cite{Duan2004}. The qubit information is encoded in the polarization of the photons. A polarizing beam splitter (PBS) directs the $h$-polarized component to the spin system and the $v$-polarized component to a mirror M. The $j$'th pulse needs to be incident twice to implement the gate. (b) Schematic setup of the active, error-proof optical Bell state analyzer of Ref.~\cite{Witthaut2012}. The qubit information is assumed encoded in the path of the photons (dual rail) and two spin systems are needed. The figures are reproduced with permission (a): Ref.~\cite{Duan2004},2004 American Physical Society and (b): Ref.~\cite{Witthaut2012}, 2012, IOP Publishing. }
\label{fig:figure1}
\end{figure}

An alternative strategy is to create strong optical non-linearities by coupling spin systems to optical resonators or waveguides. A controlled-phase gate (CZ-gate) can be realized between two photons by sequential scattering off a cavity or a waveguide coupled to a three-level system~\cite{Duan2004} (see Fig.~\ref{fig:figure1}(a)). Together with single photon Hadamard gates and detectors this enables a Bell state analyzer. 

The basic principle can be understood by considering the scattering of a single photon from a single sided cavity strongly coupled to a three-level atomic system. Assume that the atom is prepared in some arbitrary superposition of two ground states $\ket{g}$ and $\ket{s}$. The cavity field couples state $\ket{s}$ to an excited level $\ket{e}$ with single photon Rabi frequency $g$, while the other ground state $\ket{g}$ is uncoupled. In the absence of intra-cavity losses, the annihilation operator describing the scattered light will be~\cite{Anders2003}
\begin{equation}
\hat{a}_{\text{out}}=\frac{-1+4C\hat{N}_{s}}{1+4C
\hat{N}_{s}}\hat{a}_{\text{in}},
\end{equation}
assuming resonant (both with the cavity and the atomic transition) input light described by the annihilation operator $\hat{a}_{\text{in}}$. Here $C=\abs{g}^2/(\gamma \kappa)$ is the cooperativity where $\kappa$ is the intensity decay rate of the cavity field. Spontaneous emission from the excited level is assumed to be described by a Lindblad operator $\hat{L}=\sqrt{\gamma}\ketbra{s}{e}$ with $\gamma$ being the spontaneous decay rate of the excited level. The quantity $\hat{N}_{s}=\ketbra{s}{s}$ is the projector onto state $\ket{s}$. Thus, $\hat{N}_s=1$ if the atom is prepared in state $\ket{s}$ such that $\hat{a}_{\text{out}}\approx\hat{a}_{\text{in}}$ for $C\gg1$. If the atom is prepared in state $\ket{g}$, we have that $\hat{N}=0$ and $\hat{a}_{\text{out}}\approx-\hat{a}_{\text{in}}$. Consequently, the field experiences a $\pi$-phase shift depending on the atomic state.  

If a qubit is encoded in the horizontal/vertical polarization components of the photonic field and only horizontal polarization couples to the atomic system, the scattering will amount to a CZ-gate between the photonic qubit and an atomic qubit encoded in the ground states: an arbitrary state $\alpha_1\ket{V}\ket{g}+\alpha_2\ket{V}\ket{s}+\alpha_3\ket{H}\ket{g}+\alpha_4\ket{H}\ket{s}$ is transformed to $\alpha_1\ket{V}\ket{g}+\alpha_2\ket{V}\ket{s}+\alpha_3\ket{H}\ket{g}-\alpha_4\ket{H}\ket{s}$ up to a global phase in the limit $C\gg1$. Here $\ket{H}$ ($\ket{V}$) dones a horizontal (vertical) polarized photon. A photon-photon CZ gate can then be obtained through sequential scattering as described in Ref.~\cite{Duan2004}.

Comparing the single photon assisted linear optical bell state analyzer to the non-linearity based approach, the latter seems most promising for integrated photonics. While efficient coupling of quantum dots and color defects to nanophotonic resonators have already been demonstrated in experiments~\cite{Arcari2014,Sipahigil2016}, a number of additional requirements, however, have to be considered. In the original proposal~\cite{Duan2004}, one of the photons has to scatter off the spin system twice requiring fast optical routing and delay lines. This can be circumvented by introducing a second spin system~\cite{Witthaut2012} (see Fig.~\ref{fig:figure1}(b)). Importantly, with a second spin system the Bell-state analyzer can also be made error-proof in the sense that limited coupling efficiency only reduces the success probability, but never leads to the wrong outcome. Both proposals require control pulses and/or measurements on the spin systems and are thus examples of active Bell state analyzers.

\begin{figure} [h]
\centering
\includegraphics[width=0.45\textwidth]{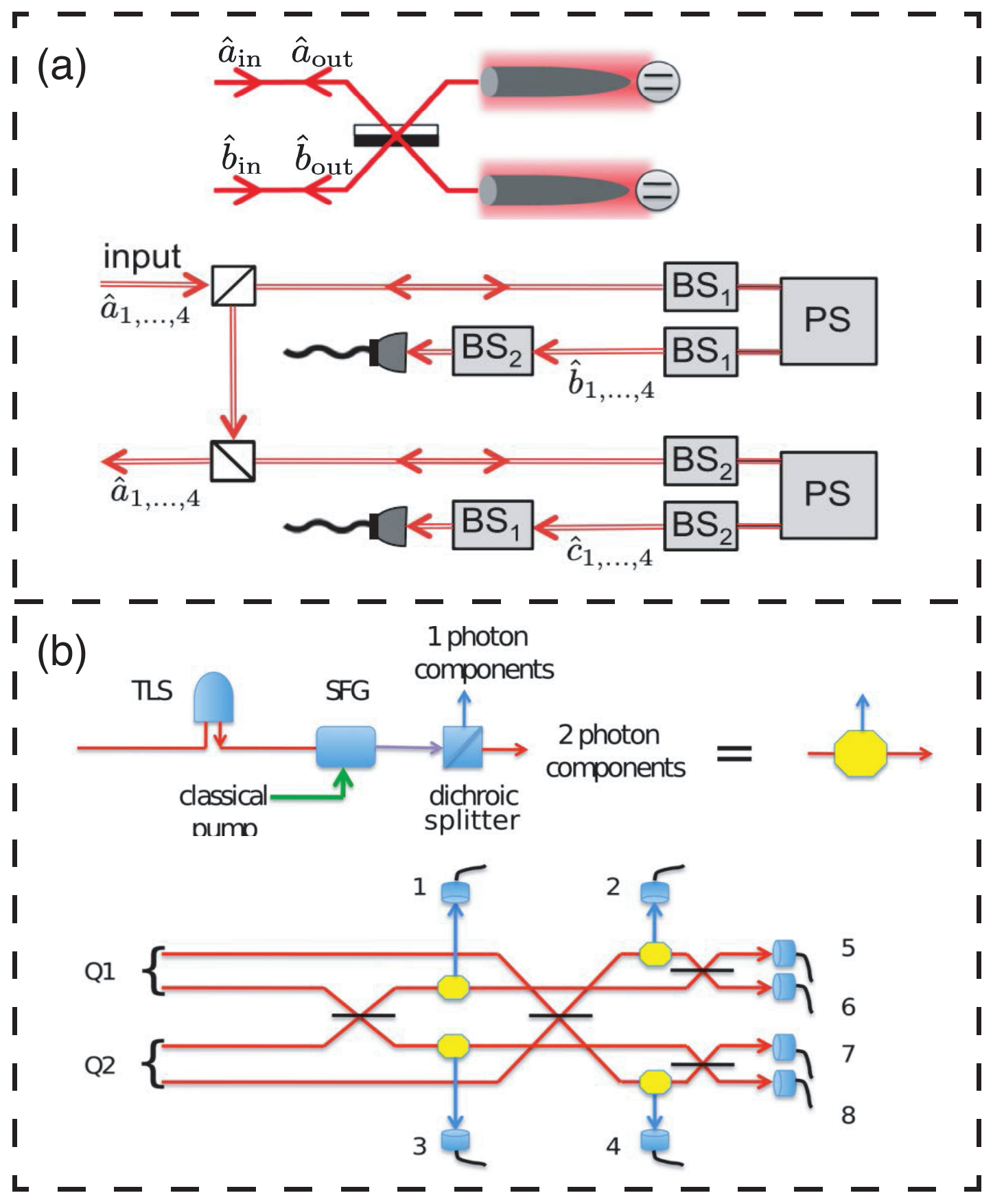}
\caption{ (a) Passive optical Bell state analyzer of Ref.~\cite{Witthaut2012} based on photon sorters. The setup for the photon sorter (PS) is shown on top while its integration into a (probabilistic) Bell state analyzer is shown in the bottom. BS${_1}$ and BS$_{2}$ are linear beam splitter arrays and the crossed squares are Faraday mirrors separating incoming and reflected modes. (b) Optical Bell state analyzer of Ref.~\cite{Ralph2015}. A deterministic photon sorter with active mode selection is shown in the top. TLS denotes a two-level emitter while SFG denotes sum frequency generation, which converts the frequency of the single-photon component that is generated in an orthogonal spectral-temporal mode to the two-photon component after the scattering process. The dichroic beam splitter subsequently separates one and two photon components. A deterministic Bell state analyzer can be constructed with four photon sorters, linear optics, and photon counting (bottom). All operations in both a) and b) can be made error proof against finite coupling efficiency, such that successful operation is heralded by clicks in the detectors and photon losses therefore only influence the success probability, not the fidelity of the operation. The figures are reproduced with permission (a): Ref.~\cite{Witthaut2012}, 2012, IOP Publishing and (b): Ref.~\cite{Ralph2015}, 2015, American Physical Society.}
\label{fig:figure2}
\end{figure}

A passive Bell state analyzer without the need for control pulses was also proposed in Ref.~\cite{Witthaut2012}. Compared to the active protocols, this protocol is based on photon sorting where the direct non-linearity associated with two photons interacting with the same optical transition is used to distinguish between zero, one, or two photon inputs (see Fig.~\ref{fig:figure2}(a)). It is, however, not possible to have a perfect and deterministic photon sorter with scattering from a single two-level spin system~\cite{Shanshan2013}. As a result, the passive Bell state analyzer is inherently probabilistic although it can be made near-deterministic through concatenated applications of it. In a similar manner, scattering from multiple two-level systems can be used to boost the fidelity of a passive and deterministic CZ gate~\cite{Brod2016}. A recent proposal also shows that a deterministic Bell state analyzer can be realized when combining photon scattering and active spectral-temporal mode selection~\cite{Ralph2015} (see Fig.~\ref{fig:figure2}(b)). The main characteristics and requirements of the different Bell state analyzers are summarized in Table~\ref{tab:table2}.

\subsection{Optical QND detection}
Quantum Non-Demolition detection (QND detection) is another desirable primitive for quantum communication. An optical QND detector makes it possible to determine the presence of a photon without destroying it. In a dual-rail encoding, where the logical states $\ket{0}$ and $\ket{1}$ correspond to a photon being in two different optical modes, this allows to repeatedly measure the qubit state, thereby increasing the measurement fidelity. A QND detection can also be used by a receiver to check whether the photon is present without disturbing the qubit information if another degree of freedom such as polarization is used to encode the qubit. As a direct application, this can be used to perform device independent quantum key distribution without the need for heralded entanglement~\cite{Vazirani2014}. 

A QND detector can be realized using the same basic mechanisms underlying the Bell state analyzers: By scattering off a three-level spin system coupled to an optical resonator or waveguide, the presence of a photon can be detected through detection of the spin levels.  QND detection of optical photons has already been demonstrated experimentally by scattering off an atom~\cite{Reiserer2013} or quantum dot~\cite{Sun2016} strongly coupled to an optical cavity. In the QND setup, the spin-system is initially prepared in a superposition $(\ket{s}+\ket{g})/\sqrt{2}$ between a ground state, $\ket{g}$, and a metastable state, $\ket{s}$. The ground state is assumed to be coupled through the waveguide or resonator mode to an excited state $\ket{e}$. The scattering of a photon on this transition will ideally result in a $\pi$ phase shift on the $\ket{g}$ state as described above. Consequently, the state of the emitter will transform into $(\ket{s}-\ket{g})/\sqrt{2}$ in the case of a scattering event while it remains in the state $(\ket{s}+\ket{g})/\sqrt{2}$ if no photon was present. Measuring the state of the emitter will thus provide information about the presence/absence of a photon. The performance of the QND detector is determined by the ratio between the coherent coupling and incoherent coupling of the spin system. For a waveguide (cavity) system, the error of the spin-based QND detector will thus be suppressed as $\sim\frac{1}{\beta_{\text{coh}}} -1$ ($\sim\frac{1}{C_{\text{coh}}}$).

\begingroup
\begin{table*} [t]
\begin{tabular}{|p{4 cm}|p{6.5 cm}|p{6.5 cm}|}
\hline
Bell state analyser & Characteristics & Requirements \\ \hline
 \begin{itemize}[noitemsep,nolistsep,leftmargin=*]\item[]Linear optics~\cite{Braunstein1995,Grice2011,Ewert2014} \end{itemize} &
 \begin{itemize}[noitemsep,nolistsep,leftmargin=*]
 \item[]Success probability:
  \begin{itemize}[noitemsep,nolistsep,leftmargin=*]
\item[-] 50\% - no auxiliary photons.
\item[-] 75\% - 4 single photon auxiliary states.
\item[-] $>\!\!75\%$ - multi-photon entangled auxiliary states.
\end{itemize}
\end{itemize} &  \begin{itemize}[noitemsep,nolistsep,leftmargin=*]
\item[-] Beam splitters.
\item[-] Single photon detectors.
\item[-] Indistinguishable photons.
\end{itemize} \\ \hline
 \begin{itemize}[noitemsep,nolistsep,leftmargin=*]\item[]Cavity CZ-gate~\cite{Duan2004} \end{itemize} & \begin{itemize}[noitemsep,nolistsep,leftmargin=*]
\item[-] Failure probability $\propto\frac{1}{C}$.
\item[-] Error from pulseshape distortion suppressed as $\sim\frac{\sigma_{\omega}}{\kappa}$.
\item[-] Error from asymmetric spontaneous emission loss $\propto\frac{1}{C^2}$.
\end{itemize} &
 \begin{itemize}[noitemsep,nolistsep,leftmargin=*]
\item[-] 3-level spin system with strong optical coupling.
\item[-] Spin control and readout.
\item[-] Single photon detection and Hadamard gates.
\item[-] Optical routing and delay.
\end{itemize} \\ \hline
 \begin{itemize}[noitemsep,nolistsep,leftmargin=*]\item[]Active Bell scheme~\cite{Witthaut2012} \end{itemize}& \begin{itemize}[noitemsep,nolistsep,leftmargin=*]
\item[-] Success probability $\sim(2\beta-1)^2$.
\item[-] Error suppressed as $\sim\left(\frac{\sigma_{\omega}}{\Gamma_{\text{1d}}+\gamma}\right)^4$.
\end{itemize} &  \begin{itemize}[noitemsep,nolistsep,leftmargin=*]
\item[-] Two 3-level spin system with strong optical coupling.
\item[-] Spin control and readout.
\item[-] Single photon detection and beam splitter.
\end{itemize} \\ \hline
 \begin{itemize}[noitemsep,nolistsep,leftmargin=*]\item[]Passive Bell scheme~\cite{Witthaut2012} \end{itemize}& \begin{itemize}[noitemsep,nolistsep,leftmargin=*]
\item[-] Maximum success probability of $\sim75\%$ for a single setup.
\item[-] Success probability $>75\%$ with many concatenated setup.
\item[-] Error from inhomogeneous coupling of emitters $\sim\left(\frac{\Delta\Gamma_{\text{1d}}}{\sigma_{\omega}}\right)^2$ , where $\Delta\Gamma_{\text{1d}}$ is the difference in waveguide decay rate of the two spins.
\end{itemize} &  \begin{itemize}[noitemsep,nolistsep,leftmargin=*]
\item[-] Eight 2-level spin system with strong optical coupling.
\item[-] Single photon detection and beam splitters.
\end{itemize} \\ \hline
 \begin{itemize}[noitemsep,nolistsep,leftmargin=*]\item[]Passive CZ scheme~\cite{Brod2016} \end{itemize}& \begin{itemize}[noitemsep,nolistsep,leftmargin=*]
 \item[]For perfect coherent scattering from $N$  two-level systems:
 \begin{itemize}[noitemsep,nolistsep,leftmargin=*]
\item[-] Error $\sim0.537N^{-1.61}$.
\item[-] Optimal width $\sim0.350N^{-0.81}\Gamma$.
\end{itemize} \end{itemize}&  \begin{itemize}[noitemsep,nolistsep,leftmargin=*]
\item[-] Multiple coupled two-level spin systems.
\item[-] Counter propagating wavepackets.
\item[-] Chiral interations.
\end{itemize} \\ \hline
\begin{itemize}[noitemsep,nolistsep,leftmargin=*]\item[]Bell scheme with active optics~\cite{Ralph2015} \end{itemize}& \begin{itemize}[noitemsep,nolistsep,leftmargin=*]
\item[-] Failure probability $\propto\frac{1-\beta}{\beta}$.
\item[-] Errors from inhomogeneous coupling of spins and non-perfect filtering.
\end{itemize} &  \begin{itemize}[noitemsep,nolistsep,leftmargin=*]
\item[-] Four 2-level spin systems with strong optical coupling.
\item[-] Single photon detection and beam splitters.
\item[-] Filtering through sum frequency generation.
\end{itemize} \\ \hline
\end{tabular}
\caption{Main characteristics and requirements of the Bell state analyzers considered in Sec.~\ref{sec:elements}. In general, Gaussian pulses with frequency width $\sigma_{\omega}$ are assumed. The parameters used to characterize the performance of the schemes are defined in Tab.~\ref{tab:table1} and perfect in/out coupling and photodetectors are assumed. \label{tab:table2}}
\end{table*}
\endgroup

\subsection{Photonic cluster generation} \label{sec:cluster}
A key advantage of solid state emitters is that they can be operated as very bright single photon sources. The fast photon emission rates in the range of GHz makes it possible to emit many photons within the typical coherence time of solid state spin states. This opens up the possibility to create multi-photon entangled states with a single emitter~\cite{schon2007} such as Greenberger-Horne-Zeilinger (GHZ)~\cite{Gheri1998} and cluster states~\cite{Lindner2009}
Depending on the specific state, different applications may be relevant. Photonic 1D cluster states and GHZ states may serve as resources for quantum enhanced metrology~\cite{Friis2017}, while 2D cluster states can serve as a resource for universal measurement-based quantum computation~\cite{Maarten2006}. Certain loss-tolerant cluster states~\cite{Varnava2006} can also be used for quantum repeater protocols as we will discuss below. 
  
It was shown in Ref.~\cite{Lindner2009} that 1D cluster states can be emitted  from a single quantum emitter. The basic mechanism behind this is the repeated excitation of the quantum emitter. Consider a quantum emitter with two ground state levels $\ket{g}$, $\ket{s}$ and two excited levels $\ket{e_g}$, $\ket{e_s}$. Assume the transitions $\ket{g}\leftrightarrow\ket{e_g}$ and $\ket{s}\leftrightarrow\ket{e_s}$ are both strongly coupled to a waveguide mode but to different polarizations (horizontal, $\ket{H}$ and vertical, $\ket{V}$). Initially, the emitter is prepared in the state $(\ket{g}+\ket{s})/\sqrt{2}$. The protocol now excites the emitter with a laser pulse to make the transformation $(\ket{g}+\ket{s})/\sqrt{2}\to(\ket{e_g}+\ket{e_s})/\sqrt{2}$. In the ideal limit, where emission is solely through the waveguide, the emitter coherently decay to the state 
\begin{equation} 
\frac{1}{\sqrt{2}}(\ket{g}\ket{H}+\ket{s}\ket{V}).
\end{equation}
Repeating the procedure $n$ times creates a GHZ state between $n$ photonic qubits~\cite{Gheri1998} and the emitter. If a rotation of the emitter is performed in between emissions, the resulting state would be a $(n+1)$-qubit 1D cluster state consisting of $n$ photonic qubits and the emitter~\cite{Lindner2009}. This protocol was recently realized in an experiment demonstrating a 1D cluster state with entanglement inferred theoretically to last up to 5 photons~\cite{Schwartz2016}. 

There have been a number of proposals for generating 2D-cluster states based on a divide-and-conquer approach where smaller (1D) states are fused together in parallel to make larger (2D) cluster states~\cite{Duan2005,Nielsen2004}. The fusion gates can be probabilistic, which makes these proposals suited for linear optics approaches. The divide-and-conquer approach enables an efficient (polynomial) scaling of resources (such as the number of single photon sources and detectors) with the cluster size despite the probabilistic operations. Nonetheless, the inherent probabilistic nature of the fusion gates can still lead to substantial overhead.

\begin{figure} [t]
\centering
\includegraphics[width=0.49\textwidth]{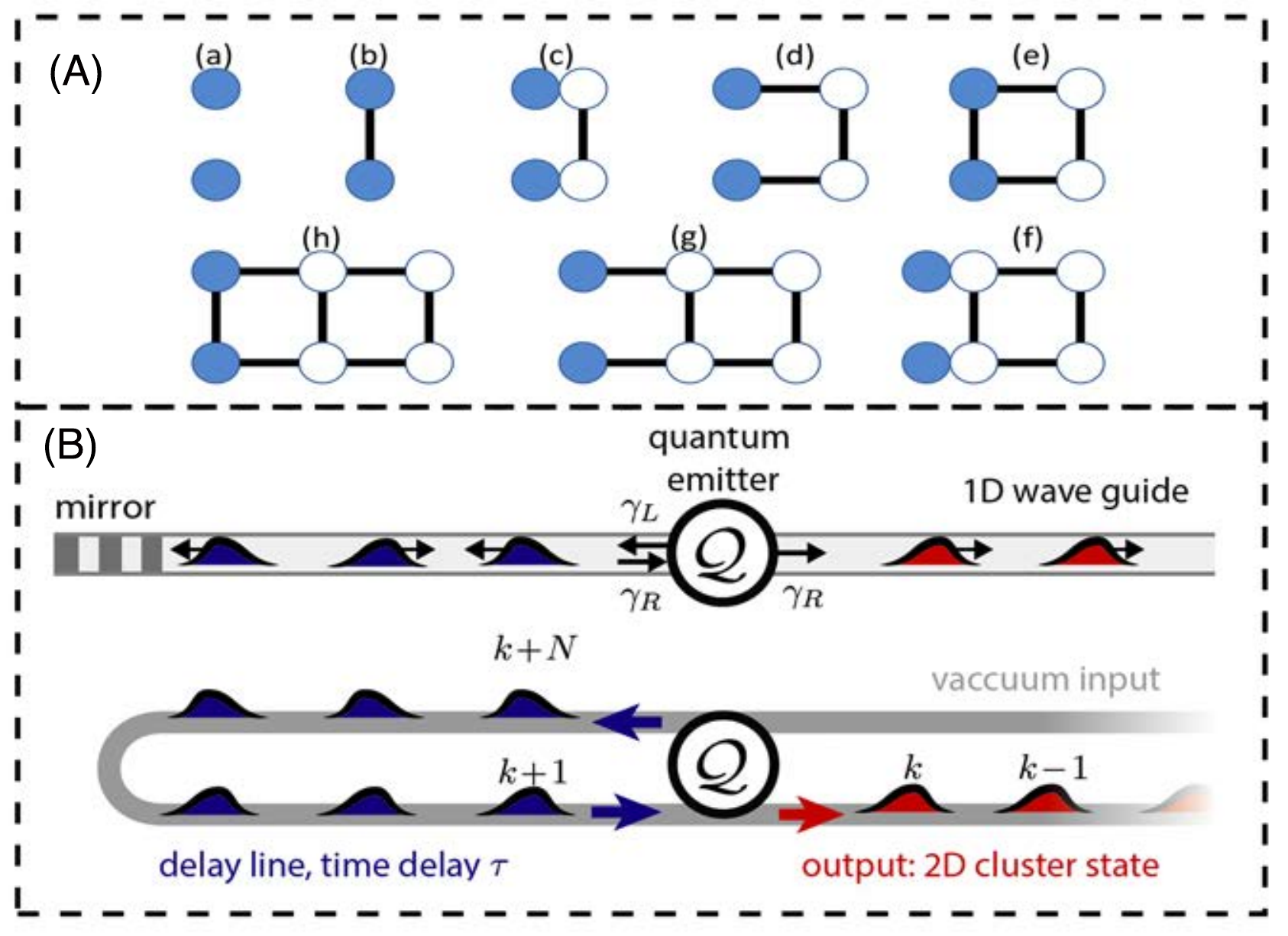}
\caption{ (A) 2D cluster generation using multiple quantum emitters~\cite{Economou2010}. Entangling gates between two emitters (blue dots) in between photon emission allows to emit connected 1D cluster strings. Following two spin rotations (a) entanglement (solid line) is created between the two emitters (b) followed by the emission of two photons (white dots) (c). This procedure is then repeated leading to the states in (d)-(h). (B) Single-emitter proposal for generating 2D photonic-cluster states~\cite{Pichler2017}. The introduced optical delay line allows the photons to interact with the emitter twice, which generates the 2D cluster state. The figures are reproduced with permission (A): Ref.~\cite{Economou2010}, 2010, American Physical Society and (B): Ref.~\cite{Pichler2017}, 2017, National Academy of Sciences. }
\label{fig:figure3}
\end{figure}

The access to non-linear quantum operations with quantum emitters opens up alternative routes to the generation of 2D cluster states. Strings of 1D cluster states being emitted from separate quantum emitters can be joined by performing entangling gates between either the photons or the emitters. The former approach can be realized using optical Bell state analyzers based on optical non-linearities as described in Sec.~\ref{sec:Bell}.  The latter approach requires direct entangling gates between the emitters in between photon emissions~\cite{Economou2010} (see Fig.~\ref{fig:figure3}(A)) with the number of quantum emitters scaling linearly with the size of the 2D cluster state. The generation protocol may however be optimized for other graph states than 2D clusters. One example is the loss-tolerant graph states considered in the all-optical repeaters of Refs.~\cite{Azuma2015,Pant2017}. An efficient scheme to generate such states has been proposed in Ref.~\cite{Buterakos2017} where the number of qubit spin systems scale logarithmically with the size of the graph state. Refs.~\cite{Economou2010} and \cite{Buterakos2017} both assume the availability of a deterministic, high-fidelity entangling gate between spin systems. For solid state systems this can be challenging to realize optically because of the inhomogeneity induced by the environment. For diamond defects (NV and SiV) coupling the electronic spin to a nuclear spin may  be used to circumvent this problem to some extent.

Another approach was suggested in Ref.~\cite{Pichler2017} where the idea is to route the photons emitted by a single emitter back to interact with the same emitter again such that they become entangled with photons emitted at  later times (see Fig.~\ref{fig:figure3}(B)). This can be done by using a delay line for the emitted photons together with suitable excitation sequence of the emitter. In this way, a 2D cluster state can be emitted in a sequential manner. Inhomogeneity due to slow drifts of the optical lines is less of a problem in this case since only a single emitter is applied, which makes this proposal very promising for solid state quantum emitters. Other cluster states, such as the loss-tolerant tree-cluster state~\cite{Varnava2006} can, in principle, be generated by performing single photon measurements together with feedforward on a 2D cluster state. More efficient generation schemes may, however, be envisioned depending on the desired cluster state and the generation scheme should, in general, be optimized based on the desired target state.

\section{Quantum repeaters} \label{sec:repeaters}

The elements described in the previous section: Bell state analyzers, QND detection, and cluster state generation may be used to overcome a key challenge for the implementation of long-distance quantum networks: photon propagation loss, which limits the distance over which quantum information can be distributed. In particular, the above mentioned resources may be used in different types of quantum-repeater protocols to enable entanglement distribution or quantum key distribution over long distances.

The goal of a quantum repeater is to reliably transmit quantum information between two distant locations in the presence of transmission loss and noise. Since the first idea of a quantum repeater was presented~\cite{Briegel1998}, numerous proposals for how to realize such devices have been formulated~\cite{Duan2001,Childress2006,Sangouard2011,Munro2012,Borregaard2015,Azuma2015,Muralidharan2016, Vinay2017}. The underlying structure of a quantum repeater has also been subject to investigation, resulting in proposals for repeater structures fundamentally different from the original one. In general, two classes of repeaters have been considered: two-way and one-way repeaters. The original repeater scheme~\cite{Briegel1998} is a two-way implementation where information has to be transmitted in both directions across the links. In contrast to this, a one-way repeater only transmits information in one direction and can therefore potentially be faster. While the two-way repeater relies on quantum memories and entanglement purification, the one-way repeater uses quantum error-correction to battle transmission loss and noise~\cite{Munro2012,Munro2015,Muralidharan2016}.

The two forms of quantum repeaters may complement each other in a quantum network depending on the application in mind. Two-way repeaters create entanglement between the stations establishing a quantum link between them to be used for e.g. distributed quantum computing~\cite{Huang2017,Wehner2018} or metrology~\cite{Komar2014,Emil2018}. To this end, they require the availability of long-term quantum memories at the repeater stations. Other applications such as quantum key distribution (QKD) do not require entanglement distribution but simply efficient transmission of a qubit from a sender to a receiver. In such cases, one-way repeaters can relax the memory requirement and boost the rate. Alternatively, one-way quantum repeaters could be used for generating entanglement between end-nodes containing large quantum memories, without the need for memories at intermediate stations. The latter is reminiscent of current classical repeaters which provide high speed connections between distant computers without having large memory and processing power.  

We will consider how the elements described in Sec.~\ref{sec:elements} may be used in both architectures. In particular, Bell state analyzers can be used both for entanglement swapping in two-way repeaters and for re-encoding information at the repeater stations for one-way repeaters. Photonic cluster states can be used as photonic memories in all-optical repeaters and QND detection may be used for device-independent quantum key distribution~\cite{Vazirani2014} with one-way repeaters.

One technical aspect is that solid-state photonic systems typically have the best optical properties at optical wavelengths, which is shorter than the telecom C-band where low-loss optical fibers exist. Frequency conversion to the telecom band is therefore necessary for long-distance quantum communication and quantum repeaters. We will not go into any details about this here, but note that recent experiments have demonstrated efficient frequency conversion of single photons emitted from quantum dots~\cite{Kambs2016} and NV centers~\cite{Dreau2018}.

\begin{figure} [t]
\centering
\includegraphics[width=0.49\textwidth]{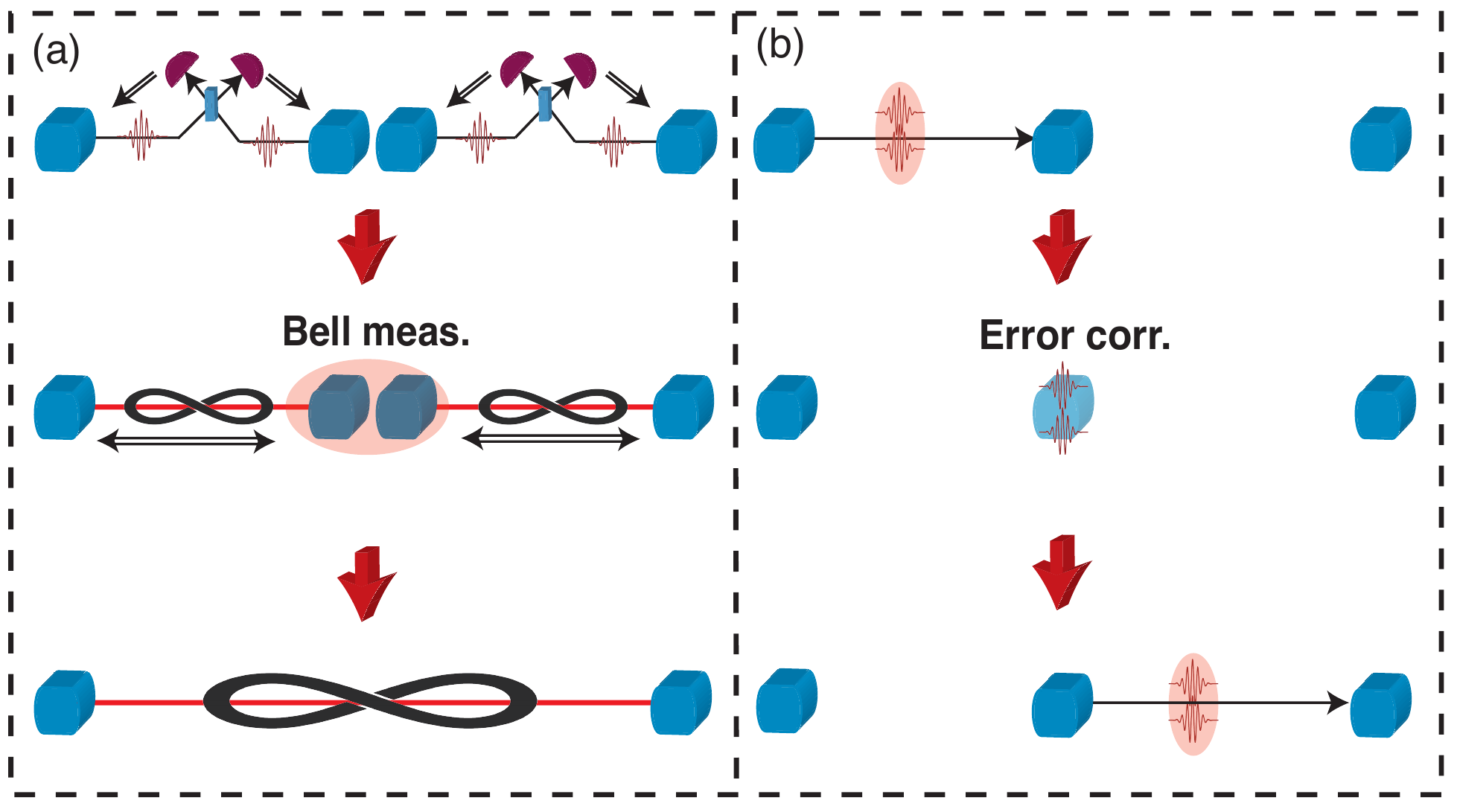}
\caption{ (a) Basic elements of a two-way quantum repeater. First, entanglement is generated between two quantum memories (indicated by blue boxes) over the elementary links in a heralded fashion. This requires the direct transmission of a quantum signal (photon wave packet) together with classical information (double arrows) signaling the success of the attempt. After neighboring links have succeeded, a Bell measurement swaps the entanglement to larger distances. Intermediate entanglement purification before the swap may be necessary and requires two-way classical communication. (b) The setup of a one-way quantum repeater. Quantum information is transmitted directly between the repeater stations in one direction. The qubit information is encoded in an error-correcting code such that transmission loss and noise can be corrected at the repeater stations. }
\label{fig:figure4}
\end{figure}

\subsection{Two-way quantum repeaters}

The general structure of a two-way quantum repeater  is shown in Fig.~\ref{fig:figure4}(a). The total distance is divided into a number of elementary links over which entanglement can be created in a heralded fashion by direct transmission of a quantum signal. There exist a number of proposals for entanglement generation schemes based on quantum emitters~\cite{Enk1997,Duan2003,Browne2003,Barrett2005}.  Two-way communication is necessary regardless of whether the entanglement is generated using a middle station~\cite{Duan2003,Browne2003,Barrett2005}, or by direct transmission over the entire link~\cite{Enk1997}. In both cases, a quantum signal has to be transmitted one way while classical information about the success of the transmission needs to be sent back to the sending station. 

Consider the scheme of Ref.~\cite{Duan2003}. In this scheme a quantum emitter strongly coupled to a cavity can coherently emit a horizontal (vertically) polarized cavity photon $\ket{H}$ ($\ket{V}$) through a decay to a ground state $\ket{0}$ ($\ket{1}$) from an exited state. The two transitions have equal coupling strengths and the emitter-cavity state after emission will thus be $(\ket{0}\ket{H}+\ket{1}\ket{V})/\sqrt{2}$. The cavity photon is now sent towards a middle station where it it is ideally combined with a photon from a similar distant system. The (uncorrelated) quantum state of the two systems is 
\begin{equation}
\frac{1}{2}(\ket{0}_1\ket{H}_1+\ket{1}_1\ket{V}_1)\otimes(\ket{0}_2\ket{H}_2+\ket{1}_2\ket{V}_2)
\end{equation}
with subscript 1 (2) denoting system 1 (2). This state can be re-written in the Bell state basis as
\begin{eqnarray} \label{eq:entgen}
&&\frac{1}{2}\big(\ket{\phi^+}_A\ket{\phi^+}_P+\ket{\phi^-}_A\ket{\phi^-}_P \nonumber \\
&&+\ket{\psi^+}_A\ket{\psi^+}_P+\ket{\psi^-}_A\ket{\psi^-}_P\big),
\end{eqnarray}
where subscript $A$ ($P$) denotes a state of the two atomic (photonic) systems. It is seen from Eq.~\ref{eq:entgen} that a Bell measurement of the photons will project the atomic systems into a Bell state and thus create entanglement between the two distant atomic systems. In Ref.~\cite{Duan2003}, the Bell measurement is 
performed with linear optics with a success probability of $\eta^2/2$, where $0\leq\eta\leq1$ is the detection probability of the photons including the transmission loss from the stations with the atomic systems to the middle station. The factor of 1/2 in the success probability is a direct consequence of the limitations of Bell state measurements using linear optics. Access to efficient Bell state measurements could thus immediately increase the entanglement generation rate by a factor of 2. 

Besides the transmission of the photons to the middle station, a classical signal thus has to be sent back to the two stations reporting the outcome of the entanglement generation attempt. Since each entanglement attempt needs two-way communication, the signaling time will limit the overall rate of the quantum repeater. Furthermore, entanglement must be created between two quantum memories. These memories are crucial both for storing quantum information during the entanglement attempt and for keeping the entangled pair until the entanglement can be swapped with a neighboring link. In this way, memories ensure that the entanglement generations in all elementary links do not have to succeed simultaneously. Once entanglement has been generated across all links, the entanglement can be extended to large distances using entanglement swapping. In the entanglement swap, one again exploits that measuring two halfs of two entangled states in the Bell basis, projects the other two halfs into an entangled states as shown in Eq.~\ref{eq:entgen}. 

The procedure discussed so far allows for entanglement distribution in the absence of errors. Real system always have errors, but these can be reduced by adding entanglement purification. In this process, two noisy entangled pairs are used to achieve an entangled pair of higher quality~\cite{Bennett1996,Deutsh1996,Kalb2017}. The process can be iterated in either a nested manner where entangled pairs of equal quality are combined or in an entanglement pumping scheme where the quality of one pair is increased by combining with a supply of lower quality pairs. The former, in general, allows for higher quality pairs, but also requires the ability to manipulate larger numbers of qubits compared to the latter approach~\cite{Briegel1998, Munro2015}. At the cost of additional communication (both classical and quantum), entanglement purification allows, in principle, to extend the entanglement to arbitrary distances in the presence of errors. The required error levels are usually at the 1\% level~\cite{Briegel1998,Childress2006,Vinay2017}. 

The entanglement generation and the memory aspect of the two-way repeater pose a number of challenges for solid state emitters. The creation of high-quality entanglement by photon interference requires indistinguishable photons from different emitters. This has been accomplished by applying electric and magnetic control fields to tune the transition frequencies of two different solid-state emitters into resonance~\cite{Delteil2017,Bernien2013}. Another question is how to realize a long-term quantum memory. The memory time should at least be on the order of the total signaling time between the two end points of the repeater. For distances of around 1000 km, this corresponds to a memory time in the ms range. A promising approach with NV centers is to use the nuclear spin of nearby carbon-13 atoms as the memory~\cite{Childress2005A,Childress2006}. Coupling the electronic spin of the NV center to the nuclear spin of the carbon-13 atom allows to transfer quantum information from the relatively short lived ($\sim \mu$s) electronic spin states to the long-lived ($\sim$ms-s) nuclear spin~\cite{Maurer2012,Kalb2017}.

\begin{figure} [t]
\centering
\includegraphics[width=0.49\textwidth]{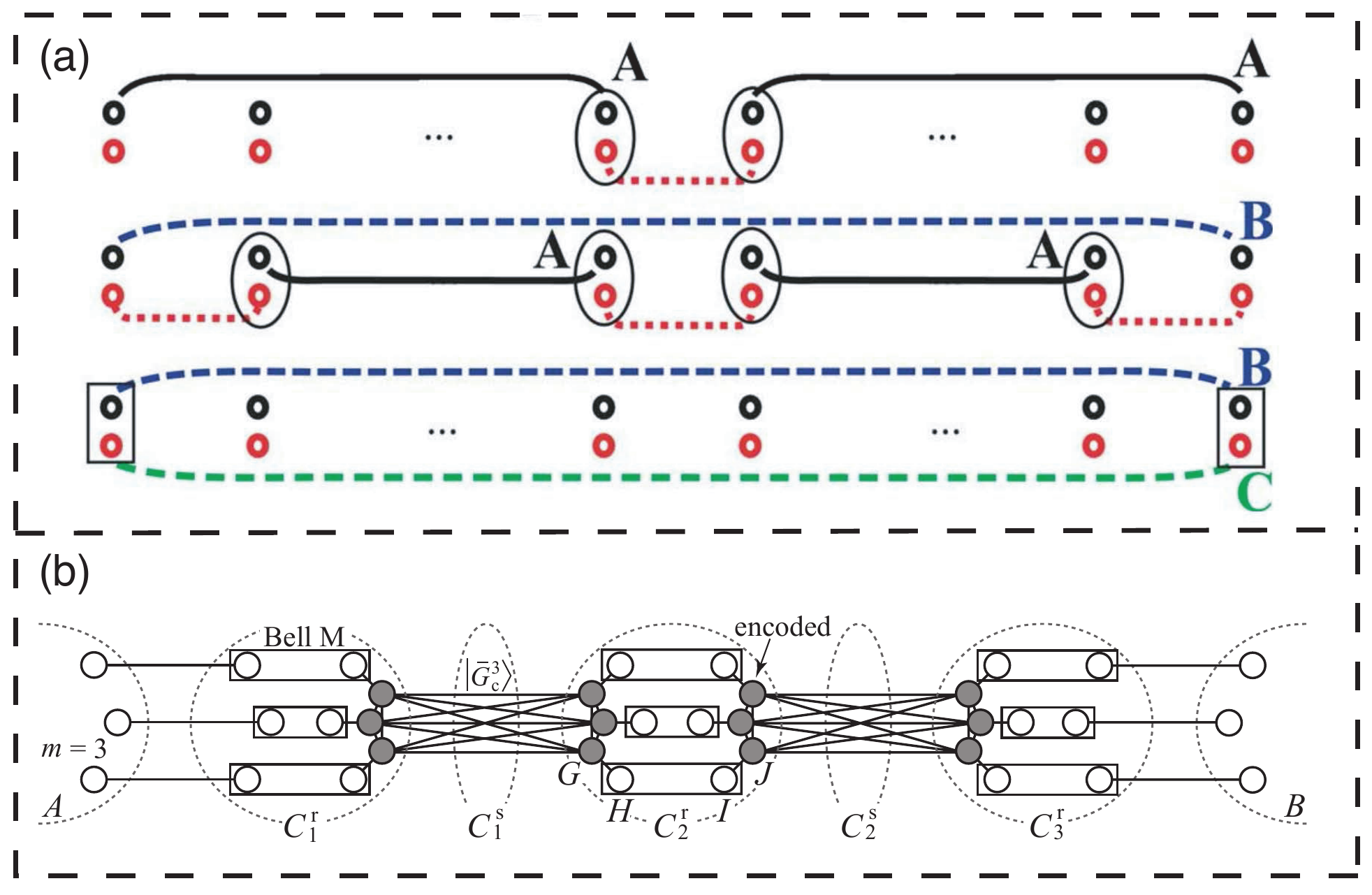}
\caption{ (a) Two-way quantum repeater with minimal resources based on NV centers in diamond with both nuclear (upper black circle) and electronic (lower red circle) spin systems. Entanglement between nodes is represented by both dashed and solid lines. Entanglement purification and swapping is represented by rectangles and ovals, respectively. (b) Sketch of an all-photonic quantum repeater. Alice (Bob) sends one half of $m$ entangled pairs to receiver nodes $C^r_1$ ($C^r_{n+1}$). At the same time all source nodes $C^s_i$ creates encoded cluster states and transmits one half of the cluster to $C^r_i$ and the other half to $C^r_{i+1}$. The receiver nodes attempt Bell measurements on the incoming photons. If at least a single Bell measurement is successful, they measure out redundant photons in the encoding to establish entanglement between adjacent receiver nodes. If no Bell measurements are successful they report a failure. At the end, the repeater nodes all announce their measurement results and entanglement between Alice and Bob is established if no node reported a failure. The figures  are reproduced with permission (a): Ref.~\cite{Childress2005A}, 2005, American Physical Society and (b): Ref.~\cite{Azuma2015}, 2015, Nature Publishing Group.}
\label{fig:figure5}
\end{figure}

Realizing a long-term quantum memory with quantum dots is not straightforward since the typical spin coherence time is at best on the order of $\mu$s~\cite{Warburton2013}. To this end, hybrid approaches have been proposed, e.g. involving coupling photons from quantum dots to an atomic ensemble~\cite{Wolters2017}. Another approach is to generate loss-tolerant photonic cluster states for creating a photonic memory for storing quantum information~\cite{Buterakos2017}. This has been considered in proposals for all-optical quantum repeaters where large loss-tolerant photonic cluster states are generated at the repeater stations~\cite{Pant2017, Azuma2015}. The cluster states are connected to signal photons that are sent to the middle stations to interfere with signal photons from the neighboring repeater station thereby entangling two neighboring cluster states. The rate of such all photonic repeaters can be boosted by also transmitting the cluster states to the middle station~\cite{Azuma2015} (see Fig.~\ref{fig:figure5}(b)).  

Optical Bell state analyzers may be employed in a number of ways in two-way quantum repeaters. For NV-based repeaters, they can be employed to perform entanglement swapping between two NV-memories by reading them out and measuring the corresponding photonic signals. For quantum repeaters based on photonic memories, they can be used in the generation of the cluster states as outlined in Sec.~\ref{sec:cluster}. For both repeater types, the Bell state analyzers can also be used for the heralded entanglement generation in the elementary links. In entanglement generation schemes with a station in the middle, a full Bell state measurement of the transmitted photons deterministically projects the corresponding memories into an entangled state. As noted above, this results in an increase of the entanglement generation rate by a factor of two compared to Bell measurements based on linear optics.

While large photonic cluster states generated by quantum dots may function as quantum memories for two-way repeaters, the NV systems arguably seem more suited for two-way repeaters due to the availability of nuclear spin memories. Notably, this also opens up the possibility of performing both entanglement purification~\cite{Kalb2017} and entanglement swapping within the same diamond in a minimum resource setup~\cite{Childress2005A,Childress2006} (see Fig.~\ref{fig:figure5}(a)). Having access to more than a single NV system will, however, allow the repeater to boost its rate through parallel entanglement generation attempts~\cite{Vinay2017}. The possibility to perform Bell measurements on different NV systems will also allow for multiplexed schemes, which can lower the memory time requirements~\cite{Collins2007}.

\subsection{One-way quantum repeaters}

Quantum dots have limited memory time but do emit photons very rapidly. They may therefore be well suited for the construction of one-way quantum repeaters. In a one-way repeater, the quantum information is encoded into an error-correcting code and transmitted from one repeater station to the next~\cite{Muralidharan2014} (see Fig.~\ref{fig:figure4}(b)). At each repeater station, the errors are corrected and the quantum information is re-encoded. This circumvents the need for long-term quantum memories since there is no waiting for a heralding signal. Consequently, the repetition rate of a one-way repeater is solely determined by the local repetition rate, i.e. how fast the errors can be corrected at the repeater stations, instead of the signaling time between repeater stations. This is, however, not true if the task of the repeater is to generate an entangled link between two remote parties. In that case, the first party has to store one part of the entangled pair while waiting for the second party to communicate that the other part was received. Quantum memories will thus still be required at the end-nodes, but the one-way repeater alleviates the requirements for quantum memories at the intermediate repeater stations. On the other hand one-way quantum repeaters are highly suited for tasks such as quantum key distribution (QKD) where secret bits of quantum information are transmitted. In this case, only classical information has to be stored at the two locations and no long-time quantum memory is required.

In combination with photonic QND detection, device-independent QKD (DI-QKD) could also be achieved in such a memory-less setting. The underlying assumption of DI-QKD is that the two parties do not trust their own measurement devices. Nonetheless, they can still obtain a secret key if they can verify that their shared correlations are strong enough to violate a Bell-inequality~\cite{Vazirani2014}. The two parties have to perform loop-hole free Bell tests in order to assure that they can share a secret key. The QND detection allows the receiving party to determine whether the transmitted qubit was lost or not before performing a measurement and can thus be used to close the detection efficiency loophole~\cite{Pearle1970,Larsson2014} in a Bell test scenario despite transmission loss.

\begin{figure} [t]
\centering
\includegraphics[width=0.49\textwidth]{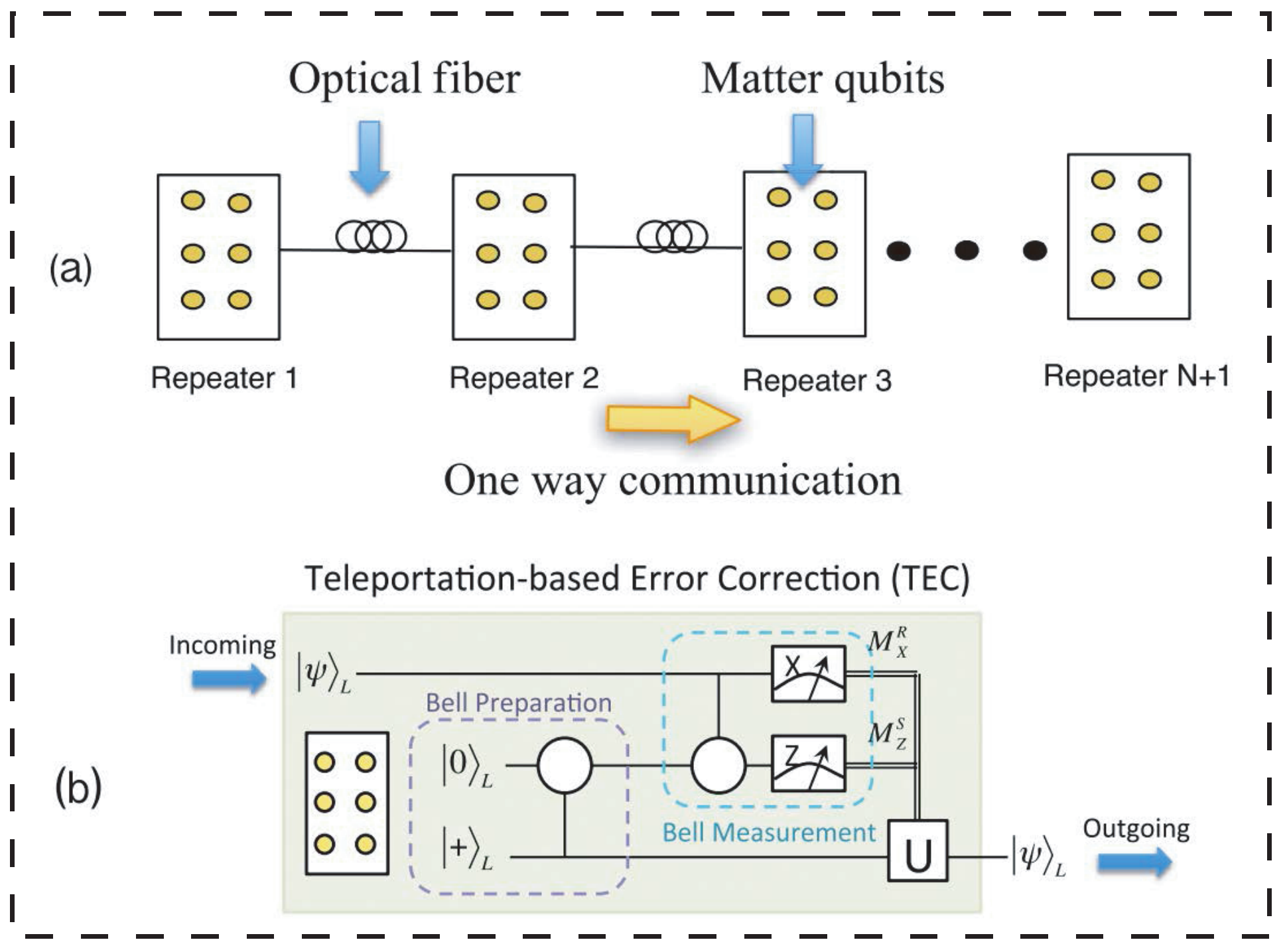}
\caption{ (a) One-way repeater of Ref.~\cite{Muralidharan2014} where the encoded photonic state is transferred to matter qubits at the repeater stations for teleportation based error correction (TEC). (b) The TEC procedure involves the generation of an encoded Bell state, which is used for a teleportation at the logical level. The teleportation operation results in an error-corrected teleported logical state. The figure is reproduced with permission from Ref.~\cite{Muralidharan2014}, 2015, American Physical Society. }
\label{fig:figure6}
\end{figure}

A number of one-way repeater schemes have been proposed based on the quantum parity code~\cite{Munro2012,Muralidharan2014,Ewert2016,Ewert2017,Lee2018}. While this code is able to correct for up to 50\% loss, the optimal spacing of repeater stations is often found to be 1-2 km corresponding to around 10\% transmission loss at telecom wavelengths~\cite{Muralidharan2014,Ewert2016,Lee2018}. The parity code involves encoding a single qubit into a multi-photon entangled state and performing teleportation-based error correction at the repeater nodes (see Fig.~\ref{fig:figure6}). The code operates with the following logical states
\begin{equation}
\ket{0}_L=\frac{1}{\sqrt{2}}(\ket{+}_L+\ket{-}_L),\quad\ket{1}=\frac{1}{\sqrt{2}}(\ket{+}_L-\ket{-}_L),
\end{equation}
with $\ket{\pm}_L=\frac{1}{\sqrt{2^n}}\left(\ket{0}^{\otimes m}\pm\ket{1}^{\otimes m}\right)^{\otimes n}$. Such a state can be generated by fusing smaller entangled states together using optical Bell analyzers. The fundamental building block of this is photonic GHZ states~\cite{Pant2017}. Spin-photon interfaces may be used both as Bell state analyzers and for the generation of photonic GHZ states as outlined in Sec.~\ref{sec:elements}. 

The teleportation-based error correction also requires Bell measurements to re-encode the quantum information at the repeater stations. At the repeater stations, a logical Bell state of the form $(\ket{0}_L\ket{0}_L+\ket{1}_L\ket{1}_L)/\sqrt{2}$ is generated and a logical Bell measurement between the incoming qubit (in general encoded as $\alpha\ket{0}_L+\beta\ket{1}_L$) and one part of the Bell pair is performed. This can be done using spin-photon gates as considered in Ref.~\cite{Muralidharan2014}. The structure of the parity code, however, also allows these Bell measurements to be done efficiently with linear optics. This was shown in Refs.~\cite{Ewert2016,Ewert2017} and used to construct a linear-optics, one-way quantum repeater without feedforward. Employing feedforward allows to reach the fundamental efficiency limit of the logical Bell measurement set by linear optics and the no-cloning theorem~\cite{Lee2018}. 

One key challenge will be to ensure that photons emitted from different emitters are indistinguishable such that high-quality Bell measurements can be performed. In particular, the parity code requires a number of Bell measurements that increases linearly with the size ($nm$) of the code in order to re-encode the information at the repeater stations. This means that for spin-based implementations hundreds of matter qubits per repeater station will be required~\cite{Muralidharan2014}.

To decrease the complexity of the quantum repeater, alternative loss-tolerant codes may be considered. In particular, matter based qudits have been considered to decrease the number of matter based quantum systems at the repeater stations~\cite{Glaudell2016,Muralidharan2017}. These works consider general Calderbank-Shor-Steane (CSS) codes in order to minimize the number of required matter based qudits at the repeater stations to on the order of $\sim10$~\cite{Glaudell2016}. As with parity code quantum repeaters, spin-photon CZ gates and QND detection are also highly desirable operations for these repeater schemes.

A significant challenge of the one-way quantum repeaters is the necessary level of noise suppression. The relevant error-correcting codes have high tolerance for loss since this is an easily detectable error. Other errors such as dephasing and depolarizing noise are harder to correct and one-way repeaters often require these to be at the $0.1\%$ level~\cite{Muralidharan2016,Muralidharan2018,Glaudell2016}.

\section{Conclusion and Discussion}
In conclusion, we have considered how the recent experimental progress towards deterministic solid state spin-photon interfaces enables implementing a number of proposals relevant for the construction of quantum networks. Specifically, we have discussed how optical Bell state analyzers, QND detectors, and photonic cluster state generation can be realized based on such hardware. We have discussed the integration of these elements into both two-way and one-way quantum repeater architectures, which are necessary to battle transmission loss and noise in future quantum networks.

As outlined in the article, the performance of the devices depends on key parameters such as coupling strength between the spin system and the optical cavity/waveguide, collection efficiency of emitted photons, spin coherence and inhomogeneous broadening. These parameters vary substantially for the range of systems currently being developed for spin-photon interfaces. Systems such as neutral atoms and diamond vacancies with nuclear spin coupling have long coherence times and thus seem very suited for applications such as two-way repeaters. In such cases, the limited collection efficiencies will limit the rate of entanglement distribution but not necessarily the fidelity in heralded entanglement generation schemes. The limited coherence time of quantum dots makes them not well suited for this purpose but their very fast photon emission rates compared to the spin coherence rates makes them very strong candidates for multi-photon entanglement sources that could be used in one-way repeaters or as elements in optical Bell state analyzers.  

A number of daunting challenges still remain in order to realize full scale quantum repeaters based on the considered systems. These include fast optical routing, efficient single-photon detectors, and the suppression of the general noise level to the $0.1\%$ level for one-way repeaters~\cite{Muralidharan2016,Muralidharan2018,Glaudell2016} and to the 1\% level for two-way repeaters~\cite{Briegel1998, Childress2006,Vinay2017}. Furthermore, high-performance entangling gate operations between emitters and conversion of optical signals to telecom wavelengths have to be developed. Nonetheless, it seems that solid state emitters are already at a level of maturity where proof-of-principle experiments of the elementary building blocks discussed in this work may be realized in the near future. Besides demonstrating interesting quantum mechanical phenomena, such experiments could also help further establish the route towards the realization of full scale quantum repeaters for future quantum networks.

\begin{acknowledgments}
We gratefully acknowledge financial support from the Danish National Research Foundation (Center of Excellence `Hy-Q', grant number DNRF139) and the European Research Council (ERC Advanced Grant `SCALE'). J. B. acknowledges support from the European Research Council (ERC Grant Agreement no. 337603) and VILLUM FONDEN via the QMATH Centre of Excellence (Grant no. 10059).
	\end{acknowledgments}
	

\begin{thebibliography}{98}%
\makeatletter
\providecommand \@ifxundefined [1]{%
 \@ifx{#1\undefined}
}%
\providecommand \@ifnum [1]{%
 \ifnum #1\expandafter \@firstoftwo
 \else \expandafter \@secondoftwo
 \fi
}%
\providecommand \@ifx [1]{%
 \ifx #1\expandafter \@firstoftwo
 \else \expandafter \@secondoftwo
 \fi
}%
\providecommand \natexlab [1]{#1}%
\providecommand \enquote  [1]{``#1''}%
\providecommand \bibnamefont  [1]{#1}%
\providecommand \bibfnamefont [1]{#1}%
\providecommand \citenamefont [1]{#1}%
\providecommand \href@noop [0]{\@secondoftwo}%
\providecommand \href [0]{\begingroup \@sanitize@url \@href}%
\providecommand \@href[1]{\@@startlink{#1}\@@href}%
\providecommand \@@href[1]{\endgroup#1\@@endlink}%
\providecommand \@sanitize@url [0]{\catcode `\\12\catcode `\$12\catcode
  `\&12\catcode `\#12\catcode `\^12\catcode `\_12\catcode `\%12\relax}%
\providecommand \@@startlink[1]{}%
\providecommand \@@endlink[0]{}%
\providecommand \url  [0]{\begingroup\@sanitize@url \@url }%
\providecommand \@url [1]{\endgroup\@href {#1}{\urlprefix }}%
\providecommand \urlprefix  [0]{URL }%
\providecommand \Eprint [0]{\href }%
\providecommand \doibase [0]{http://dx.doi.org/}%
\providecommand \selectlanguage [0]{\@gobble}%
\providecommand \bibinfo  [0]{\@secondoftwo}%
\providecommand \bibfield  [0]{\@secondoftwo}%
\providecommand \translation [1]{[#1]}%
\providecommand \BibitemOpen [0]{}%
\providecommand \bibitemStop [0]{}%
\providecommand \bibitemNoStop [0]{.\EOS\space}%
\providecommand \EOS [0]{\spacefactor3000\relax}%
\providecommand \BibitemShut  [1]{\csname bibitem#1\endcsname}%
\let\auto@bib@innerbib\@empty
\bibitem [{\citenamefont {Giovannetti}\ \emph {et~al.}(2011)\citenamefont
  {Giovannetti}, \citenamefont {Lloyd},\ and\ \citenamefont
  {Maccone}}]{Giovannetti2011}%
  \BibitemOpen
  \bibfield  {author} {\bibinfo {author} {\bibfnamefont {V.}~\bibnamefont
  {Giovannetti}}, \bibinfo {author} {\bibfnamefont {S.}~\bibnamefont {Lloyd}},
  \ and\ \bibinfo {author} {\bibfnamefont {L.}~\bibnamefont {Maccone}},\ }\href
  {http://dx.doi.org/10.1038/nphoton.2011.35} {\bibfield  {journal} {\bibinfo
  {journal} {Nature Photonics}\ }\textbf {\bibinfo {volume} {5}},\ \bibinfo
  {pages} {222} (\bibinfo {year} {2011})}\BibitemShut {NoStop}%
\bibitem [{\citenamefont {Ladd}\ \emph {et~al.}(2010)\citenamefont {Ladd},
  \citenamefont {Jelezko}, \citenamefont {Laflamme}, \citenamefont {Nakamura},
  \citenamefont {Monroe},\ and\ \citenamefont {O'Brien}}]{Ladd2010}%
  \BibitemOpen
  \bibfield  {author} {\bibinfo {author} {\bibfnamefont {T.~D.}\ \bibnamefont
  {Ladd}}, \bibinfo {author} {\bibfnamefont {F.}~\bibnamefont {Jelezko}},
  \bibinfo {author} {\bibfnamefont {R.}~\bibnamefont {Laflamme}}, \bibinfo
  {author} {\bibfnamefont {Y.}~\bibnamefont {Nakamura}}, \bibinfo {author}
  {\bibfnamefont {C.}~\bibnamefont {Monroe}}, \ and\ \bibinfo {author}
  {\bibfnamefont {J.~L.}\ \bibnamefont {O'Brien}},\ }\href
  {http://dx.doi.org/10.1038/nature08812} {\bibfield  {journal} {\bibinfo
  {journal} {Nature}\ }\textbf {\bibinfo {volume} {464}},\ \bibinfo {pages}
  {45} (\bibinfo {year} {2010})}\BibitemShut {NoStop}%
\bibitem [{\citenamefont {Gisin}\ and\ \citenamefont {Thew}(2007)}]{Gisin2007}%
  \BibitemOpen
  \bibfield  {author} {\bibinfo {author} {\bibfnamefont {N.}~\bibnamefont
  {Gisin}}\ and\ \bibinfo {author} {\bibfnamefont {R.}~\bibnamefont {Thew}},\
  }\href {http://dx.doi.org/10.1038/nphoton.2007.22} {\bibfield  {journal}
  {\bibinfo  {journal} {Nature Photonics}\ }\textbf {\bibinfo {volume} {1}},\
  \bibinfo {pages} {165} (\bibinfo {year} {2007})}\BibitemShut {NoStop}%
\bibitem [{\citenamefont {Zhang}\ \emph {et~al.}(2017)\citenamefont {Zhang},
  \citenamefont {Pagano}, \citenamefont {Hess}, \citenamefont {Kyprianidis},
  \citenamefont {Becker}, \citenamefont {Kaplan}, \citenamefont {Gorshkov},
  \citenamefont {Gong},\ and\ \citenamefont {Monroe}}]{Zhang2017}%
  \BibitemOpen
  \bibfield  {author} {\bibinfo {author} {\bibfnamefont {J.}~\bibnamefont
  {Zhang}}, \bibinfo {author} {\bibfnamefont {G.}~\bibnamefont {Pagano}},
  \bibinfo {author} {\bibfnamefont {P.~W.}\ \bibnamefont {Hess}}, \bibinfo
  {author} {\bibfnamefont {A.}~\bibnamefont {Kyprianidis}}, \bibinfo {author}
  {\bibfnamefont {P.}~\bibnamefont {Becker}}, \bibinfo {author} {\bibfnamefont
  {H.}~\bibnamefont {Kaplan}}, \bibinfo {author} {\bibfnamefont {A.~V.}\
  \bibnamefont {Gorshkov}}, \bibinfo {author} {\bibfnamefont {Z.~X.}\
  \bibnamefont {Gong}}, \ and\ \bibinfo {author} {\bibfnamefont
  {C.}~\bibnamefont {Monroe}},\ }\href {http://dx.doi.org/10.1038/nature24654}
  {\bibfield  {journal} {\bibinfo  {journal} {Nature}\ }\textbf {\bibinfo
  {volume} {551}},\ \bibinfo {pages} {601} (\bibinfo {year}
  {2017})}\BibitemShut {NoStop}%
\bibitem [{\citenamefont {Bernien}\ \emph {et~al.}(2017)\citenamefont
  {Bernien}, \citenamefont {Schwartz}, \citenamefont {Keesling}, \citenamefont
  {Levine}, \citenamefont {Omran}, \citenamefont {Pichler}, \citenamefont
  {Choi}, \citenamefont {Zibrov}, \citenamefont {Endres}, \citenamefont
  {Greiner}, \citenamefont {Vuleti{\'c}},\ and\ \citenamefont
  {Lukin}}]{Bernien2017}%
  \BibitemOpen
  \bibfield  {author} {\bibinfo {author} {\bibfnamefont {H.}~\bibnamefont
  {Bernien}}, \bibinfo {author} {\bibfnamefont {S.}~\bibnamefont {Schwartz}},
  \bibinfo {author} {\bibfnamefont {A.}~\bibnamefont {Keesling}}, \bibinfo
  {author} {\bibfnamefont {H.}~\bibnamefont {Levine}}, \bibinfo {author}
  {\bibfnamefont {A.}~\bibnamefont {Omran}}, \bibinfo {author} {\bibfnamefont
  {H.}~\bibnamefont {Pichler}}, \bibinfo {author} {\bibfnamefont
  {S.}~\bibnamefont {Choi}}, \bibinfo {author} {\bibfnamefont {A.~S.}\
  \bibnamefont {Zibrov}}, \bibinfo {author} {\bibfnamefont {M.}~\bibnamefont
  {Endres}}, \bibinfo {author} {\bibfnamefont {M.}~\bibnamefont {Greiner}},
  \bibinfo {author} {\bibfnamefont {V.}~\bibnamefont {Vuleti{\'c}}}, \ and\
  \bibinfo {author} {\bibfnamefont {M.~D.}\ \bibnamefont {Lukin}},\ }\href
  {http://dx.doi.org/10.1038/nature24622} {\bibfield  {journal} {\bibinfo
  {journal} {Nature}\ }\textbf {\bibinfo {volume} {551}},\ \bibinfo {pages}
  {579} (\bibinfo {year} {2017})}\BibitemShut {NoStop}%
\bibitem [{\citenamefont {Wendin}(2017)}]{Wendin2017}%
  \BibitemOpen
  \bibfield  {author} {\bibinfo {author} {\bibfnamefont {G.}~\bibnamefont
  {Wendin}},\ }\href {http://stacks.iop.org/0034-4885/80/i=10/a=106001}
  {\bibfield  {journal} {\bibinfo  {journal} {Reports on Progress in Physics}\
  }\textbf {\bibinfo {volume} {80}},\ \bibinfo {pages} {106001} (\bibinfo
  {year} {2017})}\BibitemShut {NoStop}%
\bibitem [{\citenamefont {Ritter}\ \emph {et~al.}(2012)\citenamefont {Ritter},
  \citenamefont {N{\"o}lleke}, \citenamefont {Hahn}, \citenamefont {Reiserer},
  \citenamefont {Neuzner}, \citenamefont {Uphoff}, \citenamefont {M{\"u}cke},
  \citenamefont {Figueroa}, \citenamefont {Bochmann},\ and\ \citenamefont
  {Rempe}}]{Ritter2012}%
  \BibitemOpen
  \bibfield  {author} {\bibinfo {author} {\bibfnamefont {S.}~\bibnamefont
  {Ritter}}, \bibinfo {author} {\bibfnamefont {C.}~\bibnamefont {N{\"o}lleke}},
  \bibinfo {author} {\bibfnamefont {C.}~\bibnamefont {Hahn}}, \bibinfo {author}
  {\bibfnamefont {A.}~\bibnamefont {Reiserer}}, \bibinfo {author}
  {\bibfnamefont {A.}~\bibnamefont {Neuzner}}, \bibinfo {author} {\bibfnamefont
  {M.}~\bibnamefont {Uphoff}}, \bibinfo {author} {\bibfnamefont
  {M.}~\bibnamefont {M{\"u}cke}}, \bibinfo {author} {\bibfnamefont
  {E.}~\bibnamefont {Figueroa}}, \bibinfo {author} {\bibfnamefont
  {J.}~\bibnamefont {Bochmann}}, \ and\ \bibinfo {author} {\bibfnamefont
  {G.}~\bibnamefont {Rempe}},\ }\href {http://dx.doi.org/10.1038/nature11023}
  {\bibfield  {journal} {\bibinfo  {journal} {Nature}\ }\textbf {\bibinfo
  {volume} {484}},\ \bibinfo {pages} {195} (\bibinfo {year}
  {2012})}\BibitemShut {NoStop}%
\bibitem [{\citenamefont {Schr\"{o}der}\ \emph {et~al.}(2016)\citenamefont
  {Schr\"{o}der}, \citenamefont {Mouradian}, \citenamefont {Zheng},
  \citenamefont {Trusheim}, \citenamefont {Walsh}, \citenamefont {Chen},
  \citenamefont {Li}, \citenamefont {Bayn},\ and\ \citenamefont
  {Englund}}]{Schroder2016}%
  \BibitemOpen
  \bibfield  {author} {\bibinfo {author} {\bibfnamefont {T.}~\bibnamefont
  {Schr\"{o}der}}, \bibinfo {author} {\bibfnamefont {S.~L.}\ \bibnamefont
  {Mouradian}}, \bibinfo {author} {\bibfnamefont {J.}~\bibnamefont {Zheng}},
  \bibinfo {author} {\bibfnamefont {M.~E.}\ \bibnamefont {Trusheim}}, \bibinfo
  {author} {\bibfnamefont {M.}~\bibnamefont {Walsh}}, \bibinfo {author}
  {\bibfnamefont {E.~H.}\ \bibnamefont {Chen}}, \bibinfo {author}
  {\bibfnamefont {L.}~\bibnamefont {Li}}, \bibinfo {author} {\bibfnamefont
  {I.}~\bibnamefont {Bayn}}, \ and\ \bibinfo {author} {\bibfnamefont
  {D.}~\bibnamefont {Englund}},\ }\href {\doibase 10.1364/JOSAB.33.000B65}
  {\bibfield  {journal} {\bibinfo  {journal} {J. Opt. Soc. Am. B}\ }\textbf
  {\bibinfo {volume} {33}},\ \bibinfo {pages} {B65} (\bibinfo {year}
  {2016})}\BibitemShut {NoStop}%
\bibitem [{\citenamefont {Lodahl}\ \emph {et~al.}(2015)\citenamefont {Lodahl},
  \citenamefont {Mahmoodian},\ and\ \citenamefont {Stobbe}}]{Lodahl2015}%
  \BibitemOpen
  \bibfield  {author} {\bibinfo {author} {\bibfnamefont {P.}~\bibnamefont
  {Lodahl}}, \bibinfo {author} {\bibfnamefont {S.}~\bibnamefont {Mahmoodian}},
  \ and\ \bibinfo {author} {\bibfnamefont {S.}~\bibnamefont {Stobbe}},\ }\href
  {\doibase 10.1103/RevModPhys.87.347} {\bibfield  {journal} {\bibinfo
  {journal} {Rev. Mod. Phys.}\ }\textbf {\bibinfo {volume} {87}},\ \bibinfo
  {pages} {347} (\bibinfo {year} {2015})}\BibitemShut {NoStop}%
\bibitem [{\citenamefont {Lodahl}(2018)}]{Lodahl2018}%
  \BibitemOpen
  \bibfield  {author} {\bibinfo {author} {\bibfnamefont {P.}~\bibnamefont
  {Lodahl}},\ }\href {http://stacks.iop.org/2058-9565/3/i=1/a=013001}
  {\bibfield  {journal} {\bibinfo  {journal} {Quantum Science and Technology}\
  }\textbf {\bibinfo {volume} {3}},\ \bibinfo {pages} {013001} (\bibinfo {year}
  {2018})}\BibitemShut {NoStop}%
\bibitem [{\citenamefont {Kimble}(1998)}]{Kimble1998}%
  \BibitemOpen
  \bibfield  {author} {\bibinfo {author} {\bibfnamefont {H.~J.}\ \bibnamefont
  {Kimble}},\ }\href {http://stacks.iop.org/1402-4896/1998/i=T76/a=019}
  {\bibfield  {journal} {\bibinfo  {journal} {Physica Scripta}\ }\textbf
  {\bibinfo {volume} {1998}},\ \bibinfo {pages} {127} (\bibinfo {year}
  {1998})}\BibitemShut {NoStop}%
\bibitem [{\citenamefont {Thompson}\ \emph {et~al.}(2013)\citenamefont
  {Thompson}, \citenamefont {Tiecke}, \citenamefont {de~Leon}, \citenamefont
  {Feist}, \citenamefont {Akimov}, \citenamefont {Gullans}, \citenamefont
  {Zibrov}, \citenamefont {Vuleti{\'c}},\ and\ \citenamefont
  {Lukin}}]{Thompson2013}%
  \BibitemOpen
  \bibfield  {author} {\bibinfo {author} {\bibfnamefont {J.~D.}\ \bibnamefont
  {Thompson}}, \bibinfo {author} {\bibfnamefont {T.~G.}\ \bibnamefont
  {Tiecke}}, \bibinfo {author} {\bibfnamefont {N.~P.}\ \bibnamefont {de~Leon}},
  \bibinfo {author} {\bibfnamefont {J.}~\bibnamefont {Feist}}, \bibinfo
  {author} {\bibfnamefont {A.~V.}\ \bibnamefont {Akimov}}, \bibinfo {author}
  {\bibfnamefont {M.}~\bibnamefont {Gullans}}, \bibinfo {author} {\bibfnamefont
  {A.~S.}\ \bibnamefont {Zibrov}}, \bibinfo {author} {\bibfnamefont
  {V.}~\bibnamefont {Vuleti{\'c}}}, \ and\ \bibinfo {author} {\bibfnamefont
  {M.~D.}\ \bibnamefont {Lukin}},\ }\href {\doibase 10.1126/science.1237125}
  {\bibfield  {journal} {\bibinfo  {journal} {Science}\ }\textbf {\bibinfo
  {volume} {340}},\ \bibinfo {pages} {1202} (\bibinfo {year}
  {2013})}\BibitemShut {NoStop}%
\bibitem [{\citenamefont {Reiserer}\ and\ \citenamefont
  {Rempe}(2015)}]{Reiserer2015}%
  \BibitemOpen
  \bibfield  {author} {\bibinfo {author} {\bibfnamefont {A.}~\bibnamefont
  {Reiserer}}\ and\ \bibinfo {author} {\bibfnamefont {G.}~\bibnamefont
  {Rempe}},\ }\href {\doibase 10.1103/RevModPhys.87.1379} {\bibfield  {journal}
  {\bibinfo  {journal} {Rev. Mod. Phys.}\ }\textbf {\bibinfo {volume} {87}},\
  \bibinfo {pages} {1379} (\bibinfo {year} {2015})}\BibitemShut {NoStop}%
\bibitem [{\citenamefont {Warburton}(2013)}]{Warburton2013}%
  \BibitemOpen
  \bibfield  {author} {\bibinfo {author} {\bibfnamefont {R.~J.}\ \bibnamefont
  {Warburton}},\ }\href {http://dx.doi.org/10.1038/nmat3585} {\bibfield
  {journal} {\bibinfo  {journal} {Nature Materials}\ }\textbf {\bibinfo
  {volume} {12}},\ \bibinfo {pages} {483} (\bibinfo {year} {2013})}\BibitemShut
  {NoStop}%
\bibitem [{\citenamefont {Senellart}\ \emph {et~al.}(2017)\citenamefont
  {Senellart}, \citenamefont {Solomon},\ and\ \citenamefont
  {White}}]{Senellart2017}%
  \BibitemOpen
  \bibfield  {author} {\bibinfo {author} {\bibfnamefont {P.}~\bibnamefont
  {Senellart}}, \bibinfo {author} {\bibfnamefont {G.}~\bibnamefont {Solomon}},
  \ and\ \bibinfo {author} {\bibfnamefont {A.}~\bibnamefont {White}},\ }\href
  {http://dx.doi.org/10.1038/nnano.2017.218} {\bibfield  {journal} {\bibinfo
  {journal} {Nature Nanotechnology}\ }\textbf {\bibinfo {volume} {12}},\
  \bibinfo {pages} {1026} (\bibinfo {year} {2017})}\BibitemShut {NoStop}%
\bibitem [{\citenamefont {Doherty}\ \emph {et~al.}(2013)\citenamefont
  {Doherty}, \citenamefont {Manson}, \citenamefont {Delaney}, \citenamefont
  {Jelezko}, \citenamefont {Wrachtrup},\ and\ \citenamefont
  {Hollenberg}}]{Doherty2013}%
  \BibitemOpen
  \bibfield  {author} {\bibinfo {author} {\bibfnamefont {M.~W.}\ \bibnamefont
  {Doherty}}, \bibinfo {author} {\bibfnamefont {N.~B.}\ \bibnamefont {Manson}},
  \bibinfo {author} {\bibfnamefont {P.}~\bibnamefont {Delaney}}, \bibinfo
  {author} {\bibfnamefont {F.}~\bibnamefont {Jelezko}}, \bibinfo {author}
  {\bibfnamefont {J.}~\bibnamefont {Wrachtrup}}, \ and\ \bibinfo {author}
  {\bibfnamefont {L.~C.}\ \bibnamefont {Hollenberg}},\ }\href {\doibase
  https://doi.org/10.1016/j.physrep.2013.02.001} {\bibfield  {journal}
  {\bibinfo  {journal} {Physics Reports}\ }\textbf {\bibinfo {volume} {528}},\
  \bibinfo {pages} {1 } (\bibinfo {year} {2013})},\ \bibinfo {note} {the
  nitrogen-vacancy colour centre in diamond}\BibitemShut {NoStop}%
\bibitem [{\citenamefont {Tiecke}\ \emph {et~al.}(2014)\citenamefont {Tiecke},
  \citenamefont {Thompson}, \citenamefont {de~Leon}, \citenamefont {Liu},
  \citenamefont {Vuleti{\'c}},\ and\ \citenamefont {Lukin}}]{Tiecke2014}%
  \BibitemOpen
  \bibfield  {author} {\bibinfo {author} {\bibfnamefont {T.~G.}\ \bibnamefont
  {Tiecke}}, \bibinfo {author} {\bibfnamefont {J.~D.}\ \bibnamefont
  {Thompson}}, \bibinfo {author} {\bibfnamefont {N.~P.}\ \bibnamefont
  {de~Leon}}, \bibinfo {author} {\bibfnamefont {L.~R.}\ \bibnamefont {Liu}},
  \bibinfo {author} {\bibfnamefont {V.}~\bibnamefont {Vuleti{\'c}}}, \ and\
  \bibinfo {author} {\bibfnamefont {M.~D.}\ \bibnamefont {Lukin}},\ }\href
  {http://dx.doi.org/10.1038/nature13188} {\bibfield  {journal} {\bibinfo
  {journal} {Nature}\ }\textbf {\bibinfo {volume} {508}},\ \bibinfo {pages}
  {241} (\bibinfo {year} {2014})}\BibitemShut {NoStop}%
\bibitem [{\citenamefont {Birnbaum}\ \emph {et~al.}(2005)\citenamefont
  {Birnbaum}, \citenamefont {Boca}, \citenamefont {Miller}, \citenamefont
  {Boozer}, \citenamefont {Northup},\ and\ \citenamefont
  {Kimble}}]{Birnbaum2005}%
  \BibitemOpen
  \bibfield  {author} {\bibinfo {author} {\bibfnamefont {K.~M.}\ \bibnamefont
  {Birnbaum}}, \bibinfo {author} {\bibfnamefont {A.}~\bibnamefont {Boca}},
  \bibinfo {author} {\bibfnamefont {R.}~\bibnamefont {Miller}}, \bibinfo
  {author} {\bibfnamefont {A.~D.}\ \bibnamefont {Boozer}}, \bibinfo {author}
  {\bibfnamefont {T.~E.}\ \bibnamefont {Northup}}, \ and\ \bibinfo {author}
  {\bibfnamefont {H.~J.}\ \bibnamefont {Kimble}},\ }\href
  {http://dx.doi.org/10.1038/nature03804} {\bibfield  {journal} {\bibinfo
  {journal} {Nature}\ }\textbf {\bibinfo {volume} {436}},\ \bibinfo {pages}
  {87} (\bibinfo {year} {2005})}\BibitemShut {NoStop}%
\bibitem [{\citenamefont {Boozer}\ \emph {et~al.}(2007)\citenamefont {Boozer},
  \citenamefont {Boca}, \citenamefont {Miller}, \citenamefont {Northup},\ and\
  \citenamefont {Kimble}}]{Boozer2007}%
  \BibitemOpen
  \bibfield  {author} {\bibinfo {author} {\bibfnamefont {A.~D.}\ \bibnamefont
  {Boozer}}, \bibinfo {author} {\bibfnamefont {A.}~\bibnamefont {Boca}},
  \bibinfo {author} {\bibfnamefont {R.}~\bibnamefont {Miller}}, \bibinfo
  {author} {\bibfnamefont {T.~E.}\ \bibnamefont {Northup}}, \ and\ \bibinfo
  {author} {\bibfnamefont {H.~J.}\ \bibnamefont {Kimble}},\ }\href {\doibase
  10.1103/PhysRevLett.98.193601} {\bibfield  {journal} {\bibinfo  {journal}
  {Phys. Rev. Lett.}\ }\textbf {\bibinfo {volume} {98}},\ \bibinfo {pages}
  {193601} (\bibinfo {year} {2007})}\BibitemShut {NoStop}%
\bibitem [{\citenamefont {Hacker}\ \emph {et~al.}(2016)\citenamefont {Hacker},
  \citenamefont {Welte}, \citenamefont {Rempe},\ and\ \citenamefont
  {Ritter}}]{Hacker2016}%
  \BibitemOpen
  \bibfield  {author} {\bibinfo {author} {\bibfnamefont {B.}~\bibnamefont
  {Hacker}}, \bibinfo {author} {\bibfnamefont {S.}~\bibnamefont {Welte}},
  \bibinfo {author} {\bibfnamefont {G.}~\bibnamefont {Rempe}}, \ and\ \bibinfo
  {author} {\bibfnamefont {S.}~\bibnamefont {Ritter}},\ }\href
  {http://dx.doi.org/10.1038/nature18592} {\bibfield  {journal} {\bibinfo
  {journal} {Nature}\ }\textbf {\bibinfo {volume} {536}},\ \bibinfo {pages}
  {193} (\bibinfo {year} {2016})}\BibitemShut {NoStop}%
\bibitem [{\citenamefont {Bernien}\ \emph {et~al.}(2013)\citenamefont
  {Bernien}, \citenamefont {Hensen}, \citenamefont {Pfaff}, \citenamefont
  {Koolstra}, \citenamefont {Blok}, \citenamefont {Robledo}, \citenamefont
  {Taminiau}, \citenamefont {Markham}, \citenamefont {Twitchen}, \citenamefont
  {Childress},\ and\ \citenamefont {Hanson}}]{Bernien2013}%
  \BibitemOpen
  \bibfield  {author} {\bibinfo {author} {\bibfnamefont {H.}~\bibnamefont
  {Bernien}}, \bibinfo {author} {\bibfnamefont {B.}~\bibnamefont {Hensen}},
  \bibinfo {author} {\bibfnamefont {W.}~\bibnamefont {Pfaff}}, \bibinfo
  {author} {\bibfnamefont {G.}~\bibnamefont {Koolstra}}, \bibinfo {author}
  {\bibfnamefont {M.~S.}\ \bibnamefont {Blok}}, \bibinfo {author}
  {\bibfnamefont {L.}~\bibnamefont {Robledo}}, \bibinfo {author} {\bibfnamefont
  {T.~H.}\ \bibnamefont {Taminiau}}, \bibinfo {author} {\bibfnamefont
  {M.}~\bibnamefont {Markham}}, \bibinfo {author} {\bibfnamefont {D.~J.}\
  \bibnamefont {Twitchen}}, \bibinfo {author} {\bibfnamefont {L.}~\bibnamefont
  {Childress}}, \ and\ \bibinfo {author} {\bibfnamefont {R.}~\bibnamefont
  {Hanson}},\ }\href {http://dx.doi.org/10.1038/nature12016} {\bibfield
  {journal} {\bibinfo  {journal} {Nature}\ }\textbf {\bibinfo {volume} {497}},\
  \bibinfo {pages} {86} (\bibinfo {year} {2013})}\BibitemShut {NoStop}%
\bibitem [{\citenamefont {Sipahigil}\ \emph {et~al.}(2016)\citenamefont
  {Sipahigil}, \citenamefont {Evans}, \citenamefont {Sukachev}, \citenamefont
  {Burek}, \citenamefont {Borregaard}, \citenamefont {Bhaskar}, \citenamefont
  {Nguyen}, \citenamefont {Pacheco}, \citenamefont {Atikian}, \citenamefont
  {Meuwly}, \citenamefont {Camacho}, \citenamefont {Jelezko}, \citenamefont
  {Bielejec}, \citenamefont {Park}, \citenamefont {Lon{\v c}ar},\ and\
  \citenamefont {Lukin}}]{Sipahigil2016}%
  \BibitemOpen
  \bibfield  {author} {\bibinfo {author} {\bibfnamefont {A.}~\bibnamefont
  {Sipahigil}}, \bibinfo {author} {\bibfnamefont {R.~E.}\ \bibnamefont
  {Evans}}, \bibinfo {author} {\bibfnamefont {D.~D.}\ \bibnamefont {Sukachev}},
  \bibinfo {author} {\bibfnamefont {M.~J.}\ \bibnamefont {Burek}}, \bibinfo
  {author} {\bibfnamefont {J.}~\bibnamefont {Borregaard}}, \bibinfo {author}
  {\bibfnamefont {M.~K.}\ \bibnamefont {Bhaskar}}, \bibinfo {author}
  {\bibfnamefont {C.~T.}\ \bibnamefont {Nguyen}}, \bibinfo {author}
  {\bibfnamefont {J.~L.}\ \bibnamefont {Pacheco}}, \bibinfo {author}
  {\bibfnamefont {H.~A.}\ \bibnamefont {Atikian}}, \bibinfo {author}
  {\bibfnamefont {C.}~\bibnamefont {Meuwly}}, \bibinfo {author} {\bibfnamefont
  {R.~M.}\ \bibnamefont {Camacho}}, \bibinfo {author} {\bibfnamefont
  {F.}~\bibnamefont {Jelezko}}, \bibinfo {author} {\bibfnamefont
  {E.}~\bibnamefont {Bielejec}}, \bibinfo {author} {\bibfnamefont
  {H.}~\bibnamefont {Park}}, \bibinfo {author} {\bibfnamefont {M.}~\bibnamefont
  {Lon{\v c}ar}}, \ and\ \bibinfo {author} {\bibfnamefont {M.~D.}\ \bibnamefont
  {Lukin}},\ }\href {\doibase 10.1126/science.aah6875} {\bibfield  {journal}
  {\bibinfo  {journal} {Science}\ }\textbf {\bibinfo {volume} {354}},\ \bibinfo
  {pages} {847} (\bibinfo {year} {2016})}\BibitemShut {NoStop}%
\bibitem [{\citenamefont {Kalb}\ \emph {et~al.}(2017)\citenamefont {Kalb},
  \citenamefont {Reiserer}, \citenamefont {Humphreys}, \citenamefont
  {Bakermans}, \citenamefont {Kamerling}, \citenamefont {Nickerson},
  \citenamefont {Benjamin}, \citenamefont {Twitchen}, \citenamefont {Markham},\
  and\ \citenamefont {Hanson}}]{Kalb2017}%
  \BibitemOpen
  \bibfield  {author} {\bibinfo {author} {\bibfnamefont {N.}~\bibnamefont
  {Kalb}}, \bibinfo {author} {\bibfnamefont {A.~A.}\ \bibnamefont {Reiserer}},
  \bibinfo {author} {\bibfnamefont {P.~C.}\ \bibnamefont {Humphreys}}, \bibinfo
  {author} {\bibfnamefont {J.~J.~W.}\ \bibnamefont {Bakermans}}, \bibinfo
  {author} {\bibfnamefont {S.~J.}\ \bibnamefont {Kamerling}}, \bibinfo {author}
  {\bibfnamefont {N.~H.}\ \bibnamefont {Nickerson}}, \bibinfo {author}
  {\bibfnamefont {S.~C.}\ \bibnamefont {Benjamin}}, \bibinfo {author}
  {\bibfnamefont {D.~J.}\ \bibnamefont {Twitchen}}, \bibinfo {author}
  {\bibfnamefont {M.}~\bibnamefont {Markham}}, \ and\ \bibinfo {author}
  {\bibfnamefont {R.}~\bibnamefont {Hanson}},\ }\href {\doibase
  10.1126/science.aan0070} {\bibfield  {journal} {\bibinfo  {journal}
  {Science}\ }\textbf {\bibinfo {volume} {356}},\ \bibinfo {pages} {928}
  (\bibinfo {year} {2017})}\BibitemShut {NoStop}%
\bibitem [{\citenamefont {Arcari}\ \emph {et~al.}(2014)\citenamefont {Arcari},
  \citenamefont {S\"ollner}, \citenamefont {Javadi}, \citenamefont
  {Lindskov~Hansen}, \citenamefont {Mahmoodian}, \citenamefont {Liu},
  \citenamefont {Thyrrestrup}, \citenamefont {Lee}, \citenamefont {Song},
  \citenamefont {Stobbe},\ and\ \citenamefont {Lodahl}}]{Arcari2014}%
  \BibitemOpen
  \bibfield  {author} {\bibinfo {author} {\bibfnamefont {M.}~\bibnamefont
  {Arcari}}, \bibinfo {author} {\bibfnamefont {I.}~\bibnamefont {S\"ollner}},
  \bibinfo {author} {\bibfnamefont {A.}~\bibnamefont {Javadi}}, \bibinfo
  {author} {\bibfnamefont {S.}~\bibnamefont {Lindskov~Hansen}}, \bibinfo
  {author} {\bibfnamefont {S.}~\bibnamefont {Mahmoodian}}, \bibinfo {author}
  {\bibfnamefont {J.}~\bibnamefont {Liu}}, \bibinfo {author} {\bibfnamefont
  {H.}~\bibnamefont {Thyrrestrup}}, \bibinfo {author} {\bibfnamefont {E.~H.}\
  \bibnamefont {Lee}}, \bibinfo {author} {\bibfnamefont {J.~D.}\ \bibnamefont
  {Song}}, \bibinfo {author} {\bibfnamefont {S.}~\bibnamefont {Stobbe}}, \ and\
  \bibinfo {author} {\bibfnamefont {P.}~\bibnamefont {Lodahl}},\ }\href
  {\doibase 10.1103/PhysRevLett.113.093603} {\bibfield  {journal} {\bibinfo
  {journal} {Phys. Rev. Lett.}\ }\textbf {\bibinfo {volume} {113}},\ \bibinfo
  {pages} {093603} (\bibinfo {year} {2014})}\BibitemShut {NoStop}%
\bibitem [{\citenamefont {Delteil}\ \emph {et~al.}(2015)\citenamefont
  {Delteil}, \citenamefont {Sun}, \citenamefont {Gao}, \citenamefont {Togan},
  \citenamefont {Faelt},\ and\ \citenamefont {Imamo{\u g}lu}}]{Delteil2015}%
  \BibitemOpen
  \bibfield  {author} {\bibinfo {author} {\bibfnamefont {A.}~\bibnamefont
  {Delteil}}, \bibinfo {author} {\bibfnamefont {Z.}~\bibnamefont {Sun}},
  \bibinfo {author} {\bibfnamefont {W.-b.}\ \bibnamefont {Gao}}, \bibinfo
  {author} {\bibfnamefont {E.}~\bibnamefont {Togan}}, \bibinfo {author}
  {\bibfnamefont {S.}~\bibnamefont {Faelt}}, \ and\ \bibinfo {author}
  {\bibfnamefont {A.}~\bibnamefont {Imamo{\u g}lu}},\ }\href
  {http://dx.doi.org/10.1038/nphys3605} {\bibfield  {journal} {\bibinfo
  {journal} {Nature Physics}\ }\textbf {\bibinfo {volume} {12}},\ \bibinfo
  {pages} {218} (\bibinfo {year} {2015})}\BibitemShut {NoStop}%
\bibitem [{\citenamefont {Kuhlmann}\ \emph {et~al.}(2015)\citenamefont
  {Kuhlmann}, \citenamefont {Prechtel}, \citenamefont {Houel}, \citenamefont
  {Ludwig}, \citenamefont {Reuter}, \citenamefont {Wieck},\ and\ \citenamefont
  {Warburton}}]{Kuhlmann2015}%
  \BibitemOpen
  \bibfield  {author} {\bibinfo {author} {\bibfnamefont {A.~V.}\ \bibnamefont
  {Kuhlmann}}, \bibinfo {author} {\bibfnamefont {J.~H.}\ \bibnamefont
  {Prechtel}}, \bibinfo {author} {\bibfnamefont {J.}~\bibnamefont {Houel}},
  \bibinfo {author} {\bibfnamefont {A.}~\bibnamefont {Ludwig}}, \bibinfo
  {author} {\bibfnamefont {D.}~\bibnamefont {Reuter}}, \bibinfo {author}
  {\bibfnamefont {A.~D.}\ \bibnamefont {Wieck}}, \ and\ \bibinfo {author}
  {\bibfnamefont {R.~J.}\ \bibnamefont {Warburton}},\ }\href
  {http://dx.doi.org/10.1038/ncomms9204} {\bibfield  {journal} {\bibinfo
  {journal} {Nature Communications}\ }\textbf {\bibinfo {volume} {6}},\
  \bibinfo {pages} {8204} (\bibinfo {year} {2015})}\BibitemShut {NoStop}%
\bibitem [{\citenamefont {De~Santis}\ \emph {et~al.}(2017)\citenamefont
  {De~Santis}, \citenamefont {Ant{\'o}n}, \citenamefont {Reznychenko},
  \citenamefont {Somaschi}, \citenamefont {Coppola}, \citenamefont {Senellart},
  \citenamefont {G{\'o}mez}, \citenamefont {Lema{\^\i}tre}, \citenamefont
  {Sagnes}, \citenamefont {White}, \citenamefont {Lanco}, \citenamefont
  {Auff{\`e}ves},\ and\ \citenamefont {Senellart}}]{De-Santis2017}%
  \BibitemOpen
  \bibfield  {author} {\bibinfo {author} {\bibfnamefont {L.}~\bibnamefont
  {De~Santis}}, \bibinfo {author} {\bibfnamefont {C.}~\bibnamefont
  {Ant{\'o}n}}, \bibinfo {author} {\bibfnamefont {B.}~\bibnamefont
  {Reznychenko}}, \bibinfo {author} {\bibfnamefont {N.}~\bibnamefont
  {Somaschi}}, \bibinfo {author} {\bibfnamefont {G.}~\bibnamefont {Coppola}},
  \bibinfo {author} {\bibfnamefont {J.}~\bibnamefont {Senellart}}, \bibinfo
  {author} {\bibfnamefont {C.}~\bibnamefont {G{\'o}mez}}, \bibinfo {author}
  {\bibfnamefont {A.}~\bibnamefont {Lema{\^\i}tre}}, \bibinfo {author}
  {\bibfnamefont {I.}~\bibnamefont {Sagnes}}, \bibinfo {author} {\bibfnamefont
  {A.~G.}\ \bibnamefont {White}}, \bibinfo {author} {\bibfnamefont
  {L.}~\bibnamefont {Lanco}}, \bibinfo {author} {\bibfnamefont
  {A.}~\bibnamefont {Auff{\`e}ves}}, \ and\ \bibinfo {author} {\bibfnamefont
  {P.}~\bibnamefont {Senellart}},\ }\href
  {http://dx.doi.org/10.1038/nnano.2017.85} {\bibfield  {journal} {\bibinfo
  {journal} {Nature Nanotechnology}\ }\textbf {\bibinfo {volume} {12}},\
  \bibinfo {pages} {663} (\bibinfo {year} {2017})}\BibitemShut {NoStop}%
\bibitem [{\citenamefont {Javadi}\ \emph {et~al.}(2018)\citenamefont {Javadi},
  \citenamefont {Ding}, \citenamefont {Appel}, \citenamefont {Mahmoodian},
  \citenamefont {L{\"o}bl}, \citenamefont {S{\"o}llner}, \citenamefont
  {Schott}, \citenamefont {Papon}, \citenamefont {Pregnolato}, \citenamefont
  {Stobbe}, \citenamefont {Midolo}, \citenamefont {Schr{\"o}der}, \citenamefont
  {Wieck}, \citenamefont {Ludwig}, \citenamefont {Warburton},\ and\
  \citenamefont {Lodahl}}]{Javadi2018}%
  \BibitemOpen
  \bibfield  {author} {\bibinfo {author} {\bibfnamefont {A.}~\bibnamefont
  {Javadi}}, \bibinfo {author} {\bibfnamefont {D.}~\bibnamefont {Ding}},
  \bibinfo {author} {\bibfnamefont {M.~H.}\ \bibnamefont {Appel}}, \bibinfo
  {author} {\bibfnamefont {S.}~\bibnamefont {Mahmoodian}}, \bibinfo {author}
  {\bibfnamefont {M.~C.}\ \bibnamefont {L{\"o}bl}}, \bibinfo {author}
  {\bibfnamefont {I.}~\bibnamefont {S{\"o}llner}}, \bibinfo {author}
  {\bibfnamefont {R.}~\bibnamefont {Schott}}, \bibinfo {author} {\bibfnamefont
  {C.}~\bibnamefont {Papon}}, \bibinfo {author} {\bibfnamefont
  {T.}~\bibnamefont {Pregnolato}}, \bibinfo {author} {\bibfnamefont
  {S.}~\bibnamefont {Stobbe}}, \bibinfo {author} {\bibfnamefont
  {L.}~\bibnamefont {Midolo}}, \bibinfo {author} {\bibfnamefont
  {T.}~\bibnamefont {Schr{\"o}der}}, \bibinfo {author} {\bibfnamefont {A.~D.}\
  \bibnamefont {Wieck}}, \bibinfo {author} {\bibfnamefont {A.}~\bibnamefont
  {Ludwig}}, \bibinfo {author} {\bibfnamefont {R.~J.}\ \bibnamefont
  {Warburton}}, \ and\ \bibinfo {author} {\bibfnamefont {P.}~\bibnamefont
  {Lodahl}},\ }\href {\doibase 10.1038/s41565-018-0091-5} {\bibfield  {journal}
  {\bibinfo  {journal} {Nature Nanotechnology}\ }\textbf {\bibinfo {volume}
  {13}},\ \bibinfo {pages} {398} (\bibinfo {year} {2018})}\BibitemShut
  {NoStop}%
\bibitem [{\citenamefont {Coles}\ \emph {et~al.}(2016)\citenamefont {Coles},
  \citenamefont {Price}, \citenamefont {Dixon}, \citenamefont {Royall},
  \citenamefont {Clarke}, \citenamefont {Kok}, \citenamefont {Skolnick},
  \citenamefont {Fox},\ and\ \citenamefont {Makhonin}}]{Coles2016}%
  \BibitemOpen
  \bibfield  {author} {\bibinfo {author} {\bibfnamefont {R.~J.}\ \bibnamefont
  {Coles}}, \bibinfo {author} {\bibfnamefont {D.~M.}\ \bibnamefont {Price}},
  \bibinfo {author} {\bibfnamefont {J.~E.}\ \bibnamefont {Dixon}}, \bibinfo
  {author} {\bibfnamefont {B.}~\bibnamefont {Royall}}, \bibinfo {author}
  {\bibfnamefont {E.}~\bibnamefont {Clarke}}, \bibinfo {author} {\bibfnamefont
  {P.}~\bibnamefont {Kok}}, \bibinfo {author} {\bibfnamefont {M.~S.}\
  \bibnamefont {Skolnick}}, \bibinfo {author} {\bibfnamefont {A.~M.}\
  \bibnamefont {Fox}}, \ and\ \bibinfo {author} {\bibfnamefont {M.~N.}\
  \bibnamefont {Makhonin}},\ }\href {https://doi.org/10.1038/ncomms11183}
  {\bibfield  {journal} {\bibinfo  {journal} {Nature Communications}\ }\textbf
  {\bibinfo {volume} {7}},\ \bibinfo {pages} {11183 EP } (\bibinfo {year}
  {2016})}\BibitemShut {NoStop}%
\bibitem [{\citenamefont {Mahmoodian}\ \emph {et~al.}(2016)\citenamefont
  {Mahmoodian}, \citenamefont {Lodahl},\ and\ \citenamefont
  {S\o{}rensen}}]{Mahmoodian2016}%
  \BibitemOpen
  \bibfield  {author} {\bibinfo {author} {\bibfnamefont {S.}~\bibnamefont
  {Mahmoodian}}, \bibinfo {author} {\bibfnamefont {P.}~\bibnamefont {Lodahl}},
  \ and\ \bibinfo {author} {\bibfnamefont {A.~S.}\ \bibnamefont
  {S\o{}rensen}},\ }\href {\doibase 10.1103/PhysRevLett.117.240501} {\bibfield
  {journal} {\bibinfo  {journal} {Phys. Rev. Lett.}\ }\textbf {\bibinfo
  {volume} {117}},\ \bibinfo {pages} {240501} (\bibinfo {year}
  {2016})}\BibitemShut {NoStop}%
\bibitem [{\citenamefont {Lodahl}\ \emph {et~al.}(2017)\citenamefont {Lodahl},
  \citenamefont {Mahmoodian}, \citenamefont {Stobbe}, \citenamefont
  {Rauschenbeutel}, \citenamefont {Schneeweiss}, \citenamefont {Volz},
  \citenamefont {Pichler},\ and\ \citenamefont {Zoller}}]{Lodahl2017}%
  \BibitemOpen
  \bibfield  {author} {\bibinfo {author} {\bibfnamefont {P.}~\bibnamefont
  {Lodahl}}, \bibinfo {author} {\bibfnamefont {S.}~\bibnamefont {Mahmoodian}},
  \bibinfo {author} {\bibfnamefont {S.}~\bibnamefont {Stobbe}}, \bibinfo
  {author} {\bibfnamefont {A.}~\bibnamefont {Rauschenbeutel}}, \bibinfo
  {author} {\bibfnamefont {P.}~\bibnamefont {Schneeweiss}}, \bibinfo {author}
  {\bibfnamefont {J.}~\bibnamefont {Volz}}, \bibinfo {author} {\bibfnamefont
  {H.}~\bibnamefont {Pichler}}, \ and\ \bibinfo {author} {\bibfnamefont
  {P.}~\bibnamefont {Zoller}},\ }\href {http://dx.doi.org/10.1038/nature21037}
  {\bibfield  {journal} {\bibinfo  {journal} {Nature}\ }\textbf {\bibinfo
  {volume} {541}},\ \bibinfo {pages} {473} (\bibinfo {year}
  {2017})}\BibitemShut {NoStop}%
\bibitem [{\citenamefont {Gorshkov}\ \emph {et~al.}(2007)\citenamefont
  {Gorshkov}, \citenamefont {Andr\'e}, \citenamefont {Fleischhauer},
  \citenamefont {S\o{}rensen},\ and\ \citenamefont {Lukin}}]{Gorshkov2007}%
  \BibitemOpen
  \bibfield  {author} {\bibinfo {author} {\bibfnamefont {A.~V.}\ \bibnamefont
  {Gorshkov}}, \bibinfo {author} {\bibfnamefont {A.}~\bibnamefont {Andr\'e}},
  \bibinfo {author} {\bibfnamefont {M.}~\bibnamefont {Fleischhauer}}, \bibinfo
  {author} {\bibfnamefont {A.~S.}\ \bibnamefont {S\o{}rensen}}, \ and\ \bibinfo
  {author} {\bibfnamefont {M.~D.}\ \bibnamefont {Lukin}},\ }\href {\doibase
  10.1103/PhysRevLett.98.123601} {\bibfield  {journal} {\bibinfo  {journal}
  {Phys. Rev. Lett.}\ }\textbf {\bibinfo {volume} {98}},\ \bibinfo {pages}
  {123601} (\bibinfo {year} {2007})}\BibitemShut {NoStop}%
\bibitem [{\citenamefont {Chang}\ \emph {et~al.}(2007)\citenamefont {Chang},
  \citenamefont {S{\o}rensen}, \citenamefont {Demler},\ and\ \citenamefont
  {Lukin}}]{Chang2007}%
  \BibitemOpen
  \bibfield  {author} {\bibinfo {author} {\bibfnamefont {D.~E.}\ \bibnamefont
  {Chang}}, \bibinfo {author} {\bibfnamefont {A.~S.}\ \bibnamefont
  {S{\o}rensen}}, \bibinfo {author} {\bibfnamefont {E.~A.}\ \bibnamefont
  {Demler}}, \ and\ \bibinfo {author} {\bibfnamefont {M.~D.}\ \bibnamefont
  {Lukin}},\ }\href {http://dx.doi.org/10.1038/nphys708} {\bibfield  {journal}
  {\bibinfo  {journal} {Nature Physics}\ }\textbf {\bibinfo {volume} {3}},\
  \bibinfo {pages} {807} (\bibinfo {year} {2007})}\BibitemShut {NoStop}%
\bibitem [{\citenamefont {L\"utkenhaus}\ \emph {et~al.}(1999)\citenamefont
  {L\"utkenhaus}, \citenamefont {Calsamiglia},\ and\ \citenamefont
  {Suominen}}]{Lutkenhaus1999}%
  \BibitemOpen
  \bibfield  {author} {\bibinfo {author} {\bibfnamefont {N.}~\bibnamefont
  {L\"utkenhaus}}, \bibinfo {author} {\bibfnamefont {J.}~\bibnamefont
  {Calsamiglia}}, \ and\ \bibinfo {author} {\bibfnamefont {K.-A.}\ \bibnamefont
  {Suominen}},\ }\href {\doibase 10.1103/PhysRevA.59.3295} {\bibfield
  {journal} {\bibinfo  {journal} {Phys. Rev. A}\ }\textbf {\bibinfo {volume}
  {59}},\ \bibinfo {pages} {3295} (\bibinfo {year} {1999})}\BibitemShut
  {NoStop}%
\bibitem [{\citenamefont {Weinfurter}(1994)}]{Weinfurter1994}%
  \BibitemOpen
  \bibfield  {author} {\bibinfo {author} {\bibfnamefont {H.}~\bibnamefont
  {Weinfurter}},\ }\href {http://stacks.iop.org/0295-5075/25/i=8/a=001}
  {\bibfield  {journal} {\bibinfo  {journal} {EPL (Europhysics Letters)}\
  }\textbf {\bibinfo {volume} {25}},\ \bibinfo {pages} {559} (\bibinfo {year}
  {1994})}\BibitemShut {NoStop}%
\bibitem [{\citenamefont {Braunstein}\ and\ \citenamefont
  {Mann}(1995)}]{Braunstein1995}%
  \BibitemOpen
  \bibfield  {author} {\bibinfo {author} {\bibfnamefont {S.~L.}\ \bibnamefont
  {Braunstein}}\ and\ \bibinfo {author} {\bibfnamefont {A.}~\bibnamefont
  {Mann}},\ }\href {\doibase 10.1103/PhysRevA.51.R1727} {\bibfield  {journal}
  {\bibinfo  {journal} {Phys. Rev. A}\ }\textbf {\bibinfo {volume} {51}},\
  \bibinfo {pages} {R1727} (\bibinfo {year} {1995})}\BibitemShut {NoStop}%
\bibitem [{\citenamefont {Michler}\ \emph {et~al.}(1996)\citenamefont
  {Michler}, \citenamefont {Mattle}, \citenamefont {Weinfurter},\ and\
  \citenamefont {Zeilinger}}]{Michler1996}%
  \BibitemOpen
  \bibfield  {author} {\bibinfo {author} {\bibfnamefont {M.}~\bibnamefont
  {Michler}}, \bibinfo {author} {\bibfnamefont {K.}~\bibnamefont {Mattle}},
  \bibinfo {author} {\bibfnamefont {H.}~\bibnamefont {Weinfurter}}, \ and\
  \bibinfo {author} {\bibfnamefont {A.}~\bibnamefont {Zeilinger}},\ }\href
  {\doibase 10.1103/PhysRevA.53.R1209} {\bibfield  {journal} {\bibinfo
  {journal} {Phys. Rev. A}\ }\textbf {\bibinfo {volume} {53}},\ \bibinfo
  {pages} {R1209} (\bibinfo {year} {1996})}\BibitemShut {NoStop}%
\bibitem [{\citenamefont {Calsamiglia}\ and\ \citenamefont
  {L{\"u}tkenhaus}(2001)}]{Calsamiglia2001}%
  \BibitemOpen
  \bibfield  {author} {\bibinfo {author} {\bibfnamefont {J.}~\bibnamefont
  {Calsamiglia}}\ and\ \bibinfo {author} {\bibfnamefont {N.}~\bibnamefont
  {L{\"u}tkenhaus}},\ }\href {\doibase 10.1007/s003400000484} {\bibfield
  {journal} {\bibinfo  {journal} {Applied Physics B}\ }\textbf {\bibinfo
  {volume} {72}},\ \bibinfo {pages} {67} (\bibinfo {year} {2001})}\BibitemShut
  {NoStop}%
\bibitem [{\citenamefont {Grice}(2011)}]{Grice2011}%
  \BibitemOpen
  \bibfield  {author} {\bibinfo {author} {\bibfnamefont {W.~P.}\ \bibnamefont
  {Grice}},\ }\href {\doibase 10.1103/PhysRevA.84.042331} {\bibfield  {journal}
  {\bibinfo  {journal} {Phys. Rev. A}\ }\textbf {\bibinfo {volume} {84}},\
  \bibinfo {pages} {042331} (\bibinfo {year} {2011})}\BibitemShut {NoStop}%
\bibitem [{\citenamefont {Ewert}\ and\ \citenamefont {van
  Loock}(2014)}]{Ewert2014}%
  \BibitemOpen
  \bibfield  {author} {\bibinfo {author} {\bibfnamefont {F.}~\bibnamefont
  {Ewert}}\ and\ \bibinfo {author} {\bibfnamefont {P.}~\bibnamefont {van
  Loock}},\ }\href {\doibase 10.1103/PhysRevLett.113.140403} {\bibfield
  {journal} {\bibinfo  {journal} {Phys. Rev. Lett.}\ }\textbf {\bibinfo
  {volume} {113}},\ \bibinfo {pages} {140403} (\bibinfo {year}
  {2014})}\BibitemShut {NoStop}%
\bibitem [{\citenamefont {Duan}\ and\ \citenamefont {Kimble}(2004)}]{Duan2004}%
  \BibitemOpen
  \bibfield  {author} {\bibinfo {author} {\bibfnamefont {L.-M.}\ \bibnamefont
  {Duan}}\ and\ \bibinfo {author} {\bibfnamefont {H.~J.}\ \bibnamefont
  {Kimble}},\ }\href {\doibase 10.1103/PhysRevLett.92.127902} {\bibfield
  {journal} {\bibinfo  {journal} {Phys. Rev. Lett.}\ }\textbf {\bibinfo
  {volume} {92}},\ \bibinfo {pages} {127902} (\bibinfo {year}
  {2004})}\BibitemShut {NoStop}%
\bibitem [{\citenamefont {S\o{}rensen}\ and\ \citenamefont
  {M\o{}lmer}(2003)}]{Anders2003}%
  \BibitemOpen
  \bibfield  {author} {\bibinfo {author} {\bibfnamefont {A.~S.}\ \bibnamefont
  {S\o{}rensen}}\ and\ \bibinfo {author} {\bibfnamefont {K.}~\bibnamefont
  {M\o{}lmer}},\ }\href {\doibase 10.1103/PhysRevLett.90.127903} {\bibfield
  {journal} {\bibinfo  {journal} {Phys. Rev. Lett.}\ }\textbf {\bibinfo
  {volume} {90}},\ \bibinfo {pages} {127903} (\bibinfo {year}
  {2003})}\BibitemShut {NoStop}%
\bibitem [{\citenamefont {Witthaut}\ \emph {et~al.}(2012)\citenamefont
  {Witthaut}, \citenamefont {Lukin},\ and\ \citenamefont
  {S{\o}rensen}}]{Witthaut2012}%
  \BibitemOpen
  \bibfield  {author} {\bibinfo {author} {\bibfnamefont {D.}~\bibnamefont
  {Witthaut}}, \bibinfo {author} {\bibfnamefont {M.~D.}\ \bibnamefont {Lukin}},
  \ and\ \bibinfo {author} {\bibfnamefont {A.~S.}\ \bibnamefont
  {S{\o}rensen}},\ }\href {http://stacks.iop.org/0295-5075/97/i=5/a=50007}
  {\bibfield  {journal} {\bibinfo  {journal} {EPL (Europhysics Letters)}\
  }\textbf {\bibinfo {volume} {97}},\ \bibinfo {pages} {50007} (\bibinfo {year}
  {2012})}\BibitemShut {NoStop}%
\bibitem [{\citenamefont {Xu}\ \emph {et~al.}(2013)\citenamefont {Xu},
  \citenamefont {Rephaeli},\ and\ \citenamefont {Fan}}]{Shanshan2013}%
  \BibitemOpen
  \bibfield  {author} {\bibinfo {author} {\bibfnamefont {S.}~\bibnamefont
  {Xu}}, \bibinfo {author} {\bibfnamefont {E.}~\bibnamefont {Rephaeli}}, \ and\
  \bibinfo {author} {\bibfnamefont {S.}~\bibnamefont {Fan}},\ }\href {\doibase
  10.1103/PhysRevLett.111.223602} {\bibfield  {journal} {\bibinfo  {journal}
  {Phys. Rev. Lett.}\ }\textbf {\bibinfo {volume} {111}},\ \bibinfo {pages}
  {223602} (\bibinfo {year} {2013})}\BibitemShut {NoStop}%
\bibitem [{\citenamefont {Brod}\ and\ \citenamefont {Combes}(2016)}]{Brod2016}%
  \BibitemOpen
  \bibfield  {author} {\bibinfo {author} {\bibfnamefont {D.~J.}\ \bibnamefont
  {Brod}}\ and\ \bibinfo {author} {\bibfnamefont {J.}~\bibnamefont {Combes}},\
  }\href {\doibase 10.1103/PhysRevLett.117.080502} {\bibfield  {journal}
  {\bibinfo  {journal} {Phys. Rev. Lett.}\ }\textbf {\bibinfo {volume} {117}},\
  \bibinfo {pages} {080502} (\bibinfo {year} {2016})}\BibitemShut {NoStop}%
\bibitem [{\citenamefont {Ralph}\ \emph {et~al.}(2015)\citenamefont {Ralph},
  \citenamefont {S\"ollner}, \citenamefont {Mahmoodian}, \citenamefont
  {White},\ and\ \citenamefont {Lodahl}}]{Ralph2015}%
  \BibitemOpen
  \bibfield  {author} {\bibinfo {author} {\bibfnamefont {T.~C.}\ \bibnamefont
  {Ralph}}, \bibinfo {author} {\bibfnamefont {I.}~\bibnamefont {S\"ollner}},
  \bibinfo {author} {\bibfnamefont {S.}~\bibnamefont {Mahmoodian}}, \bibinfo
  {author} {\bibfnamefont {A.~G.}\ \bibnamefont {White}}, \ and\ \bibinfo
  {author} {\bibfnamefont {P.}~\bibnamefont {Lodahl}},\ }\href {\doibase
  10.1103/PhysRevLett.114.173603} {\bibfield  {journal} {\bibinfo  {journal}
  {Phys. Rev. Lett.}\ }\textbf {\bibinfo {volume} {114}},\ \bibinfo {pages}
  {173603} (\bibinfo {year} {2015})}\BibitemShut {NoStop}%
\bibitem [{\citenamefont {Vazirani}\ and\ \citenamefont
  {Vidick}(2014)}]{Vazirani2014}%
  \BibitemOpen
  \bibfield  {author} {\bibinfo {author} {\bibfnamefont {U.}~\bibnamefont
  {Vazirani}}\ and\ \bibinfo {author} {\bibfnamefont {T.}~\bibnamefont
  {Vidick}},\ }\href {\doibase 10.1103/PhysRevLett.113.140501} {\bibfield
  {journal} {\bibinfo  {journal} {Phys. Rev. Lett.}\ }\textbf {\bibinfo
  {volume} {113}},\ \bibinfo {pages} {140501} (\bibinfo {year}
  {2014})}\BibitemShut {NoStop}%
\bibitem [{\citenamefont {Reiserer}\ \emph {et~al.}(2013)\citenamefont
  {Reiserer}, \citenamefont {Ritter},\ and\ \citenamefont
  {Rempe}}]{Reiserer2013}%
  \BibitemOpen
  \bibfield  {author} {\bibinfo {author} {\bibfnamefont {A.}~\bibnamefont
  {Reiserer}}, \bibinfo {author} {\bibfnamefont {S.}~\bibnamefont {Ritter}}, \
  and\ \bibinfo {author} {\bibfnamefont {G.}~\bibnamefont {Rempe}},\ }\href
  {\doibase 10.1126/science.1246164} {\bibfield  {journal} {\bibinfo  {journal}
  {Science}\ }\textbf {\bibinfo {volume} {342}},\ \bibinfo {pages} {1349}
  (\bibinfo {year} {2013})}\BibitemShut {NoStop}%
\bibitem [{\citenamefont {Sun}\ \emph {et~al.}(2016)\citenamefont {Sun},
  \citenamefont {Kim}, \citenamefont {Solomon},\ and\ \citenamefont
  {Waks}}]{Sun2016}%
  \BibitemOpen
  \bibfield  {author} {\bibinfo {author} {\bibfnamefont {S.}~\bibnamefont
  {Sun}}, \bibinfo {author} {\bibfnamefont {H.}~\bibnamefont {Kim}}, \bibinfo
  {author} {\bibfnamefont {G.~S.}\ \bibnamefont {Solomon}}, \ and\ \bibinfo
  {author} {\bibfnamefont {E.}~\bibnamefont {Waks}},\ }\href
  {http://dx.doi.org/10.1038/nnano.2015.334} {\bibfield  {journal} {\bibinfo
  {journal} {Nature Nanotechnology}\ }\textbf {\bibinfo {volume} {11}},\
  \bibinfo {pages} {539} (\bibinfo {year} {2016})}\BibitemShut {NoStop}%
\bibitem [{\citenamefont {Sch\"on}\ \emph {et~al.}(2007)\citenamefont
  {Sch\"on}, \citenamefont {Hammerer}, \citenamefont {Wolf}, \citenamefont
  {Cirac},\ and\ \citenamefont {Solano}}]{schon2007}%
  \BibitemOpen
  \bibfield  {author} {\bibinfo {author} {\bibfnamefont {C.}~\bibnamefont
  {Sch\"on}}, \bibinfo {author} {\bibfnamefont {K.}~\bibnamefont {Hammerer}},
  \bibinfo {author} {\bibfnamefont {M.~M.}\ \bibnamefont {Wolf}}, \bibinfo
  {author} {\bibfnamefont {J.~I.}\ \bibnamefont {Cirac}}, \ and\ \bibinfo
  {author} {\bibfnamefont {E.}~\bibnamefont {Solano}},\ }\href {\doibase
  10.1103/PhysRevA.75.032311} {\bibfield  {journal} {\bibinfo  {journal} {Phys.
  Rev. A}\ }\textbf {\bibinfo {volume} {75}},\ \bibinfo {pages} {032311}
  (\bibinfo {year} {2007})}\BibitemShut {NoStop}%
\bibitem [{\citenamefont {Gheri}\ \emph {et~al.}(1998)\citenamefont {Gheri},
  \citenamefont {Saavedra}, \citenamefont {T\"orm\"a}, \citenamefont {Cirac},\
  and\ \citenamefont {Zoller}}]{Gheri1998}%
  \BibitemOpen
  \bibfield  {author} {\bibinfo {author} {\bibfnamefont {K.~M.}\ \bibnamefont
  {Gheri}}, \bibinfo {author} {\bibfnamefont {C.}~\bibnamefont {Saavedra}},
  \bibinfo {author} {\bibfnamefont {P.}~\bibnamefont {T\"orm\"a}}, \bibinfo
  {author} {\bibfnamefont {J.~I.}\ \bibnamefont {Cirac}}, \ and\ \bibinfo
  {author} {\bibfnamefont {P.}~\bibnamefont {Zoller}},\ }\href {\doibase
  10.1103/PhysRevA.58.R2627} {\bibfield  {journal} {\bibinfo  {journal} {Phys.
  Rev. A}\ }\textbf {\bibinfo {volume} {58}},\ \bibinfo {pages} {R2627}
  (\bibinfo {year} {1998})}\BibitemShut {NoStop}%
\bibitem [{\citenamefont {Lindner}\ and\ \citenamefont
  {Rudolph}(2009)}]{Lindner2009}%
  \BibitemOpen
  \bibfield  {author} {\bibinfo {author} {\bibfnamefont {N.~H.}\ \bibnamefont
  {Lindner}}\ and\ \bibinfo {author} {\bibfnamefont {T.}~\bibnamefont
  {Rudolph}},\ }\href {\doibase 10.1103/PhysRevLett.103.113602} {\bibfield
  {journal} {\bibinfo  {journal} {Phys. Rev. Lett.}\ }\textbf {\bibinfo
  {volume} {103}},\ \bibinfo {pages} {113602} (\bibinfo {year}
  {2009})}\BibitemShut {NoStop}%
\bibitem [{\citenamefont {Friis}\ \emph {et~al.}(2017)\citenamefont {Friis},
  \citenamefont {Orsucci}, \citenamefont {Skotiniotis}, \citenamefont
  {Sekatski}, \citenamefont {Dunjko}, \citenamefont {Briegel},\ and\
  \citenamefont {D{\"u}r}}]{Friis2017}%
  \BibitemOpen
  \bibfield  {author} {\bibinfo {author} {\bibfnamefont {N.}~\bibnamefont
  {Friis}}, \bibinfo {author} {\bibfnamefont {D.}~\bibnamefont {Orsucci}},
  \bibinfo {author} {\bibfnamefont {M.}~\bibnamefont {Skotiniotis}}, \bibinfo
  {author} {\bibfnamefont {P.}~\bibnamefont {Sekatski}}, \bibinfo {author}
  {\bibfnamefont {V.}~\bibnamefont {Dunjko}}, \bibinfo {author} {\bibfnamefont
  {H.~J.}\ \bibnamefont {Briegel}}, \ and\ \bibinfo {author} {\bibfnamefont
  {W.}~\bibnamefont {D{\"u}r}},\ }\href
  {http://stacks.iop.org/1367-2630/19/i=6/a=063044} {\bibfield  {journal}
  {\bibinfo  {journal} {New Journal of Physics}\ }\textbf {\bibinfo {volume}
  {19}},\ \bibinfo {pages} {063044} (\bibinfo {year} {2017})}\BibitemShut
  {NoStop}%
\bibitem [{\citenamefont {Van~den Nest}\ \emph {et~al.}(2006)\citenamefont
  {Van~den Nest}, \citenamefont {Miyake}, \citenamefont {D\"ur},\ and\
  \citenamefont {Briegel}}]{Maarten2006}%
  \BibitemOpen
  \bibfield  {author} {\bibinfo {author} {\bibfnamefont {M.}~\bibnamefont
  {Van~den Nest}}, \bibinfo {author} {\bibfnamefont {A.}~\bibnamefont
  {Miyake}}, \bibinfo {author} {\bibfnamefont {W.}~\bibnamefont {D\"ur}}, \
  and\ \bibinfo {author} {\bibfnamefont {H.~J.}\ \bibnamefont {Briegel}},\
  }\href {\doibase 10.1103/PhysRevLett.97.150504} {\bibfield  {journal}
  {\bibinfo  {journal} {Phys. Rev. Lett.}\ }\textbf {\bibinfo {volume} {97}},\
  \bibinfo {pages} {150504} (\bibinfo {year} {2006})}\BibitemShut {NoStop}%
\bibitem [{\citenamefont {Varnava}\ \emph {et~al.}(2006)\citenamefont
  {Varnava}, \citenamefont {Browne},\ and\ \citenamefont
  {Rudolph}}]{Varnava2006}%
  \BibitemOpen
  \bibfield  {author} {\bibinfo {author} {\bibfnamefont {M.}~\bibnamefont
  {Varnava}}, \bibinfo {author} {\bibfnamefont {D.~E.}\ \bibnamefont {Browne}},
  \ and\ \bibinfo {author} {\bibfnamefont {T.}~\bibnamefont {Rudolph}},\ }\href
  {\doibase 10.1103/PhysRevLett.97.120501} {\bibfield  {journal} {\bibinfo
  {journal} {Phys. Rev. Lett.}\ }\textbf {\bibinfo {volume} {97}},\ \bibinfo
  {pages} {120501} (\bibinfo {year} {2006})}\BibitemShut {NoStop}%
\bibitem [{\citenamefont {Schwartz}\ \emph {et~al.}(2016)\citenamefont
  {Schwartz}, \citenamefont {Cogan}, \citenamefont {Schmidgall}, \citenamefont
  {Don}, \citenamefont {Gantz}, \citenamefont {Kenneth}, \citenamefont
  {Lindner},\ and\ \citenamefont {Gershoni}}]{Schwartz2016}%
  \BibitemOpen
  \bibfield  {author} {\bibinfo {author} {\bibfnamefont {I.}~\bibnamefont
  {Schwartz}}, \bibinfo {author} {\bibfnamefont {D.}~\bibnamefont {Cogan}},
  \bibinfo {author} {\bibfnamefont {E.~R.}\ \bibnamefont {Schmidgall}},
  \bibinfo {author} {\bibfnamefont {Y.}~\bibnamefont {Don}}, \bibinfo {author}
  {\bibfnamefont {L.}~\bibnamefont {Gantz}}, \bibinfo {author} {\bibfnamefont
  {O.}~\bibnamefont {Kenneth}}, \bibinfo {author} {\bibfnamefont {N.~H.}\
  \bibnamefont {Lindner}}, \ and\ \bibinfo {author} {\bibfnamefont
  {D.}~\bibnamefont {Gershoni}},\ }\href {\doibase 10.1126/science.aah4758}
  {\bibfield  {journal} {\bibinfo  {journal} {Science}\ }\textbf {\bibinfo
  {volume} {354}},\ \bibinfo {pages} {434} (\bibinfo {year}
  {2016})}\BibitemShut {NoStop}%
\bibitem [{\citenamefont {Duan}\ and\ \citenamefont
  {Raussendorf}(2005)}]{Duan2005}%
  \BibitemOpen
  \bibfield  {author} {\bibinfo {author} {\bibfnamefont {L.-M.}\ \bibnamefont
  {Duan}}\ and\ \bibinfo {author} {\bibfnamefont {R.}~\bibnamefont
  {Raussendorf}},\ }\href {\doibase 10.1103/PhysRevLett.95.080503} {\bibfield
  {journal} {\bibinfo  {journal} {Phys. Rev. Lett.}\ }\textbf {\bibinfo
  {volume} {95}},\ \bibinfo {pages} {080503} (\bibinfo {year}
  {2005})}\BibitemShut {NoStop}%
\bibitem [{\citenamefont {Nielsen}(2004)}]{Nielsen2004}%
  \BibitemOpen
  \bibfield  {author} {\bibinfo {author} {\bibfnamefont {M.~A.}\ \bibnamefont
  {Nielsen}},\ }\href {\doibase 10.1103/PhysRevLett.93.040503} {\bibfield
  {journal} {\bibinfo  {journal} {Phys. Rev. Lett.}\ }\textbf {\bibinfo
  {volume} {93}},\ \bibinfo {pages} {040503} (\bibinfo {year}
  {2004})}\BibitemShut {NoStop}%
\bibitem [{\citenamefont {Economou}\ \emph {et~al.}(2010)\citenamefont
  {Economou}, \citenamefont {Lindner},\ and\ \citenamefont
  {Rudolph}}]{Economou2010}%
  \BibitemOpen
  \bibfield  {author} {\bibinfo {author} {\bibfnamefont {S.~E.}\ \bibnamefont
  {Economou}}, \bibinfo {author} {\bibfnamefont {N.}~\bibnamefont {Lindner}}, \
  and\ \bibinfo {author} {\bibfnamefont {T.}~\bibnamefont {Rudolph}},\ }\href
  {\doibase 10.1103/PhysRevLett.105.093601} {\bibfield  {journal} {\bibinfo
  {journal} {Phys. Rev. Lett.}\ }\textbf {\bibinfo {volume} {105}},\ \bibinfo
  {pages} {093601} (\bibinfo {year} {2010})}\BibitemShut {NoStop}%
\bibitem [{\citenamefont {Azuma}\ \emph {et~al.}(2015)\citenamefont {Azuma},
  \citenamefont {Tamaki},\ and\ \citenamefont {Lo}}]{Azuma2015}%
  \BibitemOpen
  \bibfield  {author} {\bibinfo {author} {\bibfnamefont {K.}~\bibnamefont
  {Azuma}}, \bibinfo {author} {\bibfnamefont {K.}~\bibnamefont {Tamaki}}, \
  and\ \bibinfo {author} {\bibfnamefont {H.-K.}\ \bibnamefont {Lo}},\ }\href
  {http://dx.doi.org/10.1038/ncomms7787} {\bibfield  {journal} {\bibinfo
  {journal} {Nature Communications}\ }\textbf {\bibinfo {volume} {6}},\
  \bibinfo {pages} {6787} (\bibinfo {year} {2015})}\BibitemShut {NoStop}%
\bibitem [{\citenamefont {Pant}\ \emph {et~al.}(2017)\citenamefont {Pant},
  \citenamefont {Krovi}, \citenamefont {Englund},\ and\ \citenamefont
  {Guha}}]{Pant2017}%
  \BibitemOpen
  \bibfield  {author} {\bibinfo {author} {\bibfnamefont {M.}~\bibnamefont
  {Pant}}, \bibinfo {author} {\bibfnamefont {H.}~\bibnamefont {Krovi}},
  \bibinfo {author} {\bibfnamefont {D.}~\bibnamefont {Englund}}, \ and\
  \bibinfo {author} {\bibfnamefont {S.}~\bibnamefont {Guha}},\ }\href {\doibase
  10.1103/PhysRevA.95.012304} {\bibfield  {journal} {\bibinfo  {journal} {Phys.
  Rev. A}\ }\textbf {\bibinfo {volume} {95}},\ \bibinfo {pages} {012304}
  (\bibinfo {year} {2017})}\BibitemShut {NoStop}%
\bibitem [{\citenamefont {Buterakos}\ \emph {et~al.}(2017)\citenamefont
  {Buterakos}, \citenamefont {Barnes},\ and\ \citenamefont
  {Economou}}]{Buterakos2017}%
  \BibitemOpen
  \bibfield  {author} {\bibinfo {author} {\bibfnamefont {D.}~\bibnamefont
  {Buterakos}}, \bibinfo {author} {\bibfnamefont {E.}~\bibnamefont {Barnes}}, \
  and\ \bibinfo {author} {\bibfnamefont {S.~E.}\ \bibnamefont {Economou}},\
  }\href {\doibase 10.1103/PhysRevX.7.041023} {\bibfield  {journal} {\bibinfo
  {journal} {Phys. Rev. X}\ }\textbf {\bibinfo {volume} {7}},\ \bibinfo {pages}
  {041023} (\bibinfo {year} {2017})}\BibitemShut {NoStop}%
\bibitem [{\citenamefont {Pichler}\ \emph {et~al.}(2017)\citenamefont
  {Pichler}, \citenamefont {Choi}, \citenamefont {Zoller},\ and\ \citenamefont
  {Lukin}}]{Pichler2017}%
  \BibitemOpen
  \bibfield  {author} {\bibinfo {author} {\bibfnamefont {H.}~\bibnamefont
  {Pichler}}, \bibinfo {author} {\bibfnamefont {S.}~\bibnamefont {Choi}},
  \bibinfo {author} {\bibfnamefont {P.}~\bibnamefont {Zoller}}, \ and\ \bibinfo
  {author} {\bibfnamefont {M.~D.}\ \bibnamefont {Lukin}},\ }\href {\doibase
  10.1073/pnas.1711003114} {\bibfield  {journal} {\bibinfo  {journal}
  {Proceedings of the National Academy of Sciences}\ } (\bibinfo {year}
  {2017}),\ 10.1073/pnas.1711003114}\BibitemShut {NoStop}%
\bibitem [{\citenamefont {Briegel}\ \emph {et~al.}(1998)\citenamefont
  {Briegel}, \citenamefont {D\"ur}, \citenamefont {Cirac},\ and\ \citenamefont
  {Zoller}}]{Briegel1998}%
  \BibitemOpen
  \bibfield  {author} {\bibinfo {author} {\bibfnamefont {H.-J.}\ \bibnamefont
  {Briegel}}, \bibinfo {author} {\bibfnamefont {W.}~\bibnamefont {D\"ur}},
  \bibinfo {author} {\bibfnamefont {J.~I.}\ \bibnamefont {Cirac}}, \ and\
  \bibinfo {author} {\bibfnamefont {P.}~\bibnamefont {Zoller}},\ }\href
  {\doibase 10.1103/PhysRevLett.81.5932} {\bibfield  {journal} {\bibinfo
  {journal} {Phys. Rev. Lett.}\ }\textbf {\bibinfo {volume} {81}},\ \bibinfo
  {pages} {5932} (\bibinfo {year} {1998})}\BibitemShut {NoStop}%
\bibitem [{\citenamefont {Duan}\ \emph {et~al.}(2001)\citenamefont {Duan},
  \citenamefont {Lukin}, \citenamefont {Cirac},\ and\ \citenamefont
  {Zoller}}]{Duan2001}%
  \BibitemOpen
  \bibfield  {author} {\bibinfo {author} {\bibfnamefont {L.~M.}\ \bibnamefont
  {Duan}}, \bibinfo {author} {\bibfnamefont {M.~D.}\ \bibnamefont {Lukin}},
  \bibinfo {author} {\bibfnamefont {J.~I.}\ \bibnamefont {Cirac}}, \ and\
  \bibinfo {author} {\bibfnamefont {P.}~\bibnamefont {Zoller}},\ }\href
  {http://dx.doi.org/10.1038/35106500} {\bibfield  {journal} {\bibinfo
  {journal} {Nature}\ }\textbf {\bibinfo {volume} {414}},\ \bibinfo {pages}
  {413} (\bibinfo {year} {2001})}\BibitemShut {NoStop}%
\bibitem [{\citenamefont {Childress}\ \emph {et~al.}(2006)\citenamefont
  {Childress}, \citenamefont {Taylor}, \citenamefont {S\o{}rensen},\ and\
  \citenamefont {Lukin}}]{Childress2006}%
  \BibitemOpen
  \bibfield  {author} {\bibinfo {author} {\bibfnamefont {L.}~\bibnamefont
  {Childress}}, \bibinfo {author} {\bibfnamefont {J.~M.}\ \bibnamefont
  {Taylor}}, \bibinfo {author} {\bibfnamefont {A.~S.}\ \bibnamefont
  {S\o{}rensen}}, \ and\ \bibinfo {author} {\bibfnamefont {M.~D.}\ \bibnamefont
  {Lukin}},\ }\href {\doibase 10.1103/PhysRevLett.96.070504} {\bibfield
  {journal} {\bibinfo  {journal} {Phys. Rev. Lett.}\ }\textbf {\bibinfo
  {volume} {96}},\ \bibinfo {pages} {070504} (\bibinfo {year}
  {2006})}\BibitemShut {NoStop}%
\bibitem [{\citenamefont {Sangouard}\ \emph {et~al.}(2011)\citenamefont
  {Sangouard}, \citenamefont {Simon}, \citenamefont {de~Riedmatten},\ and\
  \citenamefont {Gisin}}]{Sangouard2011}%
  \BibitemOpen
  \bibfield  {author} {\bibinfo {author} {\bibfnamefont {N.}~\bibnamefont
  {Sangouard}}, \bibinfo {author} {\bibfnamefont {C.}~\bibnamefont {Simon}},
  \bibinfo {author} {\bibfnamefont {H.}~\bibnamefont {de~Riedmatten}}, \ and\
  \bibinfo {author} {\bibfnamefont {N.}~\bibnamefont {Gisin}},\ }\href
  {\doibase 10.1103/RevModPhys.83.33} {\bibfield  {journal} {\bibinfo
  {journal} {Rev. Mod. Phys.}\ }\textbf {\bibinfo {volume} {83}},\ \bibinfo
  {pages} {33} (\bibinfo {year} {2011})}\BibitemShut {NoStop}%
\bibitem [{\citenamefont {Munro}\ \emph {et~al.}(2012)\citenamefont {Munro},
  \citenamefont {Stephens}, \citenamefont {Devitt}, \citenamefont {Harrison},\
  and\ \citenamefont {Nemoto}}]{Munro2012}%
  \BibitemOpen
  \bibfield  {author} {\bibinfo {author} {\bibfnamefont {W.~J.}\ \bibnamefont
  {Munro}}, \bibinfo {author} {\bibfnamefont {A.~M.}\ \bibnamefont {Stephens}},
  \bibinfo {author} {\bibfnamefont {S.~J.}\ \bibnamefont {Devitt}}, \bibinfo
  {author} {\bibfnamefont {K.~A.}\ \bibnamefont {Harrison}}, \ and\ \bibinfo
  {author} {\bibfnamefont {K.}~\bibnamefont {Nemoto}},\ }\href
  {http://dx.doi.org/10.1038/nphoton.2012.243} {\bibfield  {journal} {\bibinfo
  {journal} {Nature Photonics}\ }\textbf {\bibinfo {volume} {6}},\ \bibinfo
  {pages} {777} (\bibinfo {year} {2012})}\BibitemShut {NoStop}%
\bibitem [{\citenamefont {Borregaard}\ \emph {et~al.}(2015)\citenamefont
  {Borregaard}, \citenamefont {K\'om\'ar}, \citenamefont {Kessler},
  \citenamefont {Lukin},\ and\ \citenamefont {S\o{}rensen}}]{Borregaard2015}%
  \BibitemOpen
  \bibfield  {author} {\bibinfo {author} {\bibfnamefont {J.}~\bibnamefont
  {Borregaard}}, \bibinfo {author} {\bibfnamefont {P.}~\bibnamefont
  {K\'om\'ar}}, \bibinfo {author} {\bibfnamefont {E.~M.}\ \bibnamefont
  {Kessler}}, \bibinfo {author} {\bibfnamefont {M.~D.}\ \bibnamefont {Lukin}},
  \ and\ \bibinfo {author} {\bibfnamefont {A.~S.}\ \bibnamefont
  {S\o{}rensen}},\ }\href {\doibase 10.1103/PhysRevA.92.012307} {\bibfield
  {journal} {\bibinfo  {journal} {Phys. Rev. A}\ }\textbf {\bibinfo {volume}
  {92}},\ \bibinfo {pages} {012307} (\bibinfo {year} {2015})}\BibitemShut
  {NoStop}%
\bibitem [{\citenamefont {Muralidharan}\ \emph {et~al.}(2016)\citenamefont
  {Muralidharan}, \citenamefont {Li}, \citenamefont {Kim}, \citenamefont
  {L{\"u}tkenhaus}, \citenamefont {Lukin},\ and\ \citenamefont
  {Jiang}}]{Muralidharan2016}%
  \BibitemOpen
  \bibfield  {author} {\bibinfo {author} {\bibfnamefont {S.}~\bibnamefont
  {Muralidharan}}, \bibinfo {author} {\bibfnamefont {L.}~\bibnamefont {Li}},
  \bibinfo {author} {\bibfnamefont {J.}~\bibnamefont {Kim}}, \bibinfo {author}
  {\bibfnamefont {N.}~\bibnamefont {L{\"u}tkenhaus}}, \bibinfo {author}
  {\bibfnamefont {M.~D.}\ \bibnamefont {Lukin}}, \ and\ \bibinfo {author}
  {\bibfnamefont {L.}~\bibnamefont {Jiang}},\ }\href
  {http://dx.doi.org/10.1038/srep20463} {\bibfield  {journal} {\bibinfo
  {journal} {Scientific Reports}\ }\textbf {\bibinfo {volume} {6}},\ \bibinfo
  {pages} {20463} (\bibinfo {year} {2016})}\BibitemShut {NoStop}%
\bibitem [{\citenamefont {Vinay}\ and\ \citenamefont {Kok}(2017)}]{Vinay2017}%
  \BibitemOpen
  \bibfield  {author} {\bibinfo {author} {\bibfnamefont {S.~E.}\ \bibnamefont
  {Vinay}}\ and\ \bibinfo {author} {\bibfnamefont {P.}~\bibnamefont {Kok}},\
  }\href {\doibase 10.1103/PhysRevA.95.052336} {\bibfield  {journal} {\bibinfo
  {journal} {Phys. Rev. A}\ }\textbf {\bibinfo {volume} {95}},\ \bibinfo
  {pages} {052336} (\bibinfo {year} {2017})}\BibitemShut {NoStop}%
\bibitem [{\citenamefont {Munro}\ \emph {et~al.}(2015)\citenamefont {Munro},
  \citenamefont {Azuma}, \citenamefont {Tamaki},\ and\ \citenamefont
  {Nemoto}}]{Munro2015}%
  \BibitemOpen
  \bibfield  {author} {\bibinfo {author} {\bibfnamefont {W.~J.}\ \bibnamefont
  {Munro}}, \bibinfo {author} {\bibfnamefont {K.}~\bibnamefont {Azuma}},
  \bibinfo {author} {\bibfnamefont {K.}~\bibnamefont {Tamaki}}, \ and\ \bibinfo
  {author} {\bibfnamefont {K.}~\bibnamefont {Nemoto}},\ }\href {\doibase
  10.1109/JSTQE.2015.2392076} {\bibfield  {journal} {\bibinfo  {journal} {IEEE
  Journal of Selected Topics in Quantum Electronics}\ }\textbf {\bibinfo
  {volume} {21}},\ \bibinfo {pages} {78} (\bibinfo {year} {2015})}\BibitemShut
  {NoStop}%
\bibitem [{\citenamefont {Huang}\ \emph {et~al.}(2017)\citenamefont {Huang},
  \citenamefont {Zhao}, \citenamefont {Ma}, \citenamefont {Liu}, \citenamefont
  {Su}, \citenamefont {Wang}, \citenamefont {Li}, \citenamefont {Liu},
  \citenamefont {Sanders}, \citenamefont {Lu},\ and\ \citenamefont
  {Pan}}]{Huang2017}%
  \BibitemOpen
  \bibfield  {author} {\bibinfo {author} {\bibfnamefont {H.-L.}\ \bibnamefont
  {Huang}}, \bibinfo {author} {\bibfnamefont {Q.}~\bibnamefont {Zhao}},
  \bibinfo {author} {\bibfnamefont {X.}~\bibnamefont {Ma}}, \bibinfo {author}
  {\bibfnamefont {C.}~\bibnamefont {Liu}}, \bibinfo {author} {\bibfnamefont
  {Z.-E.}\ \bibnamefont {Su}}, \bibinfo {author} {\bibfnamefont {X.-L.}\
  \bibnamefont {Wang}}, \bibinfo {author} {\bibfnamefont {L.}~\bibnamefont
  {Li}}, \bibinfo {author} {\bibfnamefont {N.-L.}\ \bibnamefont {Liu}},
  \bibinfo {author} {\bibfnamefont {B.~C.}\ \bibnamefont {Sanders}}, \bibinfo
  {author} {\bibfnamefont {C.-Y.}\ \bibnamefont {Lu}}, \ and\ \bibinfo {author}
  {\bibfnamefont {J.-W.}\ \bibnamefont {Pan}},\ }\href {\doibase
  10.1103/PhysRevLett.119.050503} {\bibfield  {journal} {\bibinfo  {journal}
  {Phys. Rev. Lett.}\ }\textbf {\bibinfo {volume} {119}},\ \bibinfo {pages}
  {050503} (\bibinfo {year} {2017})}\BibitemShut {NoStop}%
\bibitem [{\citenamefont {Wehner}\ \emph {et~al.}(2018)\citenamefont {Wehner},
  \citenamefont {Elkouss},\ and\ \citenamefont {Hanson}}]{Wehner2018}%
  \BibitemOpen
  \bibfield  {author} {\bibinfo {author} {\bibfnamefont {S.}~\bibnamefont
  {Wehner}}, \bibinfo {author} {\bibfnamefont {D.}~\bibnamefont {Elkouss}}, \
  and\ \bibinfo {author} {\bibfnamefont {R.}~\bibnamefont {Hanson}},\ }\href
  {\doibase 10.1126/science.aam9288} {\bibfield  {journal} {\bibinfo  {journal}
  {Science}\ }\textbf {\bibinfo {volume} {362}} (\bibinfo {year} {2018}),\
  10.1126/science.aam9288}\BibitemShut {NoStop}%
\bibitem [{\citenamefont {K{\'o}m{\'a}r}\ \emph {et~al.}(2014)\citenamefont
  {K{\'o}m{\'a}r}, \citenamefont {Kessler}, \citenamefont {Bishof},
  \citenamefont {Jiang}, \citenamefont {S{\o}rensen}, \citenamefont {Ye},\ and\
  \citenamefont {Lukin}}]{Komar2014}%
  \BibitemOpen
  \bibfield  {author} {\bibinfo {author} {\bibfnamefont {P.}~\bibnamefont
  {K{\'o}m{\'a}r}}, \bibinfo {author} {\bibfnamefont {E.~M.}\ \bibnamefont
  {Kessler}}, \bibinfo {author} {\bibfnamefont {M.}~\bibnamefont {Bishof}},
  \bibinfo {author} {\bibfnamefont {L.}~\bibnamefont {Jiang}}, \bibinfo
  {author} {\bibfnamefont {A.~S.}\ \bibnamefont {S{\o}rensen}}, \bibinfo
  {author} {\bibfnamefont {J.}~\bibnamefont {Ye}}, \ and\ \bibinfo {author}
  {\bibfnamefont {M.~D.}\ \bibnamefont {Lukin}},\ }\href
  {https://doi.org/10.1038/nphys3000} {\bibfield  {journal} {\bibinfo
  {journal} {Nature Physics}\ }\textbf {\bibinfo {volume} {10}},\ \bibinfo
  {pages} {582 EP } (\bibinfo {year} {2014})}\BibitemShut {NoStop}%
\bibitem [{\citenamefont {Khabiboulline}\ \emph {et~al.}(2018)\citenamefont
  {Khabiboulline}, \citenamefont {Borregaard}, \citenamefont {De~Greve},\ and\
  \citenamefont {Lukin}}]{Emil2018}%
  \BibitemOpen
  \bibfield  {author} {\bibinfo {author} {\bibfnamefont {E.~T.}\ \bibnamefont
  {Khabiboulline}}, \bibinfo {author} {\bibfnamefont {J.}~\bibnamefont
  {Borregaard}}, \bibinfo {author} {\bibfnamefont {K.}~\bibnamefont
  {De~Greve}}, \ and\ \bibinfo {author} {\bibfnamefont {M.~D.}\ \bibnamefont
  {Lukin}},\ }\href@noop {} {\bibfield  {journal} {\bibinfo  {journal}
  {arXiv:1809.01659 [quant-ph]}\ } (\bibinfo {year} {2018})}\BibitemShut
  {NoStop}%
\bibitem [{\citenamefont {Kambs}\ \emph {et~al.}(2016)\citenamefont {Kambs},
  \citenamefont {Kettler}, \citenamefont {Bock}, \citenamefont {Becker},
  \citenamefont {Arend}, \citenamefont {Lenhard}, \citenamefont {Portalupi},
  \citenamefont {Jetter}, \citenamefont {Michler},\ and\ \citenamefont
  {Becher}}]{Kambs2016}%
  \BibitemOpen
  \bibfield  {author} {\bibinfo {author} {\bibfnamefont {B.}~\bibnamefont
  {Kambs}}, \bibinfo {author} {\bibfnamefont {J.}~\bibnamefont {Kettler}},
  \bibinfo {author} {\bibfnamefont {M.}~\bibnamefont {Bock}}, \bibinfo {author}
  {\bibfnamefont {J.~N.}\ \bibnamefont {Becker}}, \bibinfo {author}
  {\bibfnamefont {C.}~\bibnamefont {Arend}}, \bibinfo {author} {\bibfnamefont
  {A.}~\bibnamefont {Lenhard}}, \bibinfo {author} {\bibfnamefont {S.~L.}\
  \bibnamefont {Portalupi}}, \bibinfo {author} {\bibfnamefont {M.}~\bibnamefont
  {Jetter}}, \bibinfo {author} {\bibfnamefont {P.}~\bibnamefont {Michler}}, \
  and\ \bibinfo {author} {\bibfnamefont {C.}~\bibnamefont {Becher}},\ }\href
  {\doibase 10.1364/OE.24.022250} {\bibfield  {journal} {\bibinfo  {journal}
  {Opt. Express}\ }\textbf {\bibinfo {volume} {24}},\ \bibinfo {pages} {22250}
  (\bibinfo {year} {2016})}\BibitemShut {NoStop}%
\bibitem [{\citenamefont {Dr\'eau}\ \emph {et~al.}(2018)\citenamefont
  {Dr\'eau}, \citenamefont {Tchebotareva}, \citenamefont {Mahdaoui},
  \citenamefont {Bonato},\ and\ \citenamefont {Hanson}}]{Dreau2018}%
  \BibitemOpen
  \bibfield  {author} {\bibinfo {author} {\bibfnamefont {A.}~\bibnamefont
  {Dr\'eau}}, \bibinfo {author} {\bibfnamefont {A.}~\bibnamefont
  {Tchebotareva}}, \bibinfo {author} {\bibfnamefont {A.~E.}\ \bibnamefont
  {Mahdaoui}}, \bibinfo {author} {\bibfnamefont {C.}~\bibnamefont {Bonato}}, \
  and\ \bibinfo {author} {\bibfnamefont {R.}~\bibnamefont {Hanson}},\ }\href
  {\doibase 10.1103/PhysRevApplied.9.064031} {\bibfield  {journal} {\bibinfo
  {journal} {Phys. Rev. Applied}\ }\textbf {\bibinfo {volume} {9}},\ \bibinfo
  {pages} {064031} (\bibinfo {year} {2018})}\BibitemShut {NoStop}%
\bibitem [{\citenamefont {van Enk}\ \emph {et~al.}(1997)\citenamefont {van
  Enk}, \citenamefont {Cirac},\ and\ \citenamefont {Zoller}}]{Enk1997}%
  \BibitemOpen
  \bibfield  {author} {\bibinfo {author} {\bibfnamefont {S.~J.}\ \bibnamefont
  {van Enk}}, \bibinfo {author} {\bibfnamefont {J.~I.}\ \bibnamefont {Cirac}},
  \ and\ \bibinfo {author} {\bibfnamefont {P.}~\bibnamefont {Zoller}},\ }\href
  {\doibase 10.1103/PhysRevLett.78.4293} {\bibfield  {journal} {\bibinfo
  {journal} {Phys. Rev. Lett.}\ }\textbf {\bibinfo {volume} {78}},\ \bibinfo
  {pages} {4293} (\bibinfo {year} {1997})}\BibitemShut {NoStop}%
\bibitem [{\citenamefont {Duan}\ and\ \citenamefont {Kimble}(2003)}]{Duan2003}%
  \BibitemOpen
  \bibfield  {author} {\bibinfo {author} {\bibfnamefont {L.-M.}\ \bibnamefont
  {Duan}}\ and\ \bibinfo {author} {\bibfnamefont {H.~J.}\ \bibnamefont
  {Kimble}},\ }\href {\doibase 10.1103/PhysRevLett.90.253601} {\bibfield
  {journal} {\bibinfo  {journal} {Phys. Rev. Lett.}\ }\textbf {\bibinfo
  {volume} {90}},\ \bibinfo {pages} {253601} (\bibinfo {year}
  {2003})}\BibitemShut {NoStop}%
\bibitem [{\citenamefont {Browne}\ \emph {et~al.}(2003)\citenamefont {Browne},
  \citenamefont {Plenio},\ and\ \citenamefont {Huelga}}]{Browne2003}%
  \BibitemOpen
  \bibfield  {author} {\bibinfo {author} {\bibfnamefont {D.~E.}\ \bibnamefont
  {Browne}}, \bibinfo {author} {\bibfnamefont {M.~B.}\ \bibnamefont {Plenio}},
  \ and\ \bibinfo {author} {\bibfnamefont {S.~F.}\ \bibnamefont {Huelga}},\
  }\href {\doibase 10.1103/PhysRevLett.91.067901} {\bibfield  {journal}
  {\bibinfo  {journal} {Phys. Rev. Lett.}\ }\textbf {\bibinfo {volume} {91}},\
  \bibinfo {pages} {067901} (\bibinfo {year} {2003})}\BibitemShut {NoStop}%
\bibitem [{\citenamefont {Barrett}\ and\ \citenamefont
  {Kok}(2005)}]{Barrett2005}%
  \BibitemOpen
  \bibfield  {author} {\bibinfo {author} {\bibfnamefont {S.~D.}\ \bibnamefont
  {Barrett}}\ and\ \bibinfo {author} {\bibfnamefont {P.}~\bibnamefont {Kok}},\
  }\href {\doibase 10.1103/PhysRevA.71.060310} {\bibfield  {journal} {\bibinfo
  {journal} {Phys. Rev. A}\ }\textbf {\bibinfo {volume} {71}},\ \bibinfo
  {pages} {060310} (\bibinfo {year} {2005})}\BibitemShut {NoStop}%
\bibitem [{\citenamefont {Bennett}\ \emph {et~al.}(1996)\citenamefont
  {Bennett}, \citenamefont {Brassard}, \citenamefont {Popescu}, \citenamefont
  {Schumacher}, \citenamefont {Smolin},\ and\ \citenamefont
  {Wootters}}]{Bennett1996}%
  \BibitemOpen
  \bibfield  {author} {\bibinfo {author} {\bibfnamefont {C.~H.}\ \bibnamefont
  {Bennett}}, \bibinfo {author} {\bibfnamefont {G.}~\bibnamefont {Brassard}},
  \bibinfo {author} {\bibfnamefont {S.}~\bibnamefont {Popescu}}, \bibinfo
  {author} {\bibfnamefont {B.}~\bibnamefont {Schumacher}}, \bibinfo {author}
  {\bibfnamefont {J.~A.}\ \bibnamefont {Smolin}}, \ and\ \bibinfo {author}
  {\bibfnamefont {W.~K.}\ \bibnamefont {Wootters}},\ }\href {\doibase
  10.1103/PhysRevLett.76.722} {\bibfield  {journal} {\bibinfo  {journal} {Phys.
  Rev. Lett.}\ }\textbf {\bibinfo {volume} {76}},\ \bibinfo {pages} {722}
  (\bibinfo {year} {1996})}\BibitemShut {NoStop}%
\bibitem [{\citenamefont {Deutsch}\ \emph {et~al.}(1996)\citenamefont
  {Deutsch}, \citenamefont {Ekert}, \citenamefont {Jozsa}, \citenamefont
  {Macchiavello}, \citenamefont {Popescu},\ and\ \citenamefont
  {Sanpera}}]{Deutsh1996}%
  \BibitemOpen
  \bibfield  {author} {\bibinfo {author} {\bibfnamefont {D.}~\bibnamefont
  {Deutsch}}, \bibinfo {author} {\bibfnamefont {A.}~\bibnamefont {Ekert}},
  \bibinfo {author} {\bibfnamefont {R.}~\bibnamefont {Jozsa}}, \bibinfo
  {author} {\bibfnamefont {C.}~\bibnamefont {Macchiavello}}, \bibinfo {author}
  {\bibfnamefont {S.}~\bibnamefont {Popescu}}, \ and\ \bibinfo {author}
  {\bibfnamefont {A.}~\bibnamefont {Sanpera}},\ }\href {\doibase
  10.1103/PhysRevLett.77.2818} {\bibfield  {journal} {\bibinfo  {journal}
  {Phys. Rev. Lett.}\ }\textbf {\bibinfo {volume} {77}},\ \bibinfo {pages}
  {2818} (\bibinfo {year} {1996})}\BibitemShut {NoStop}%
\bibitem [{\citenamefont {Delteil}\ \emph {et~al.}(2017)\citenamefont
  {Delteil}, \citenamefont {Sun}, \citenamefont {F\"alt},\ and\ \citenamefont
  {Imamo\ifmmode~\breve{g}\else \u{g}\fi{}lu}}]{Delteil2017}%
  \BibitemOpen
  \bibfield  {author} {\bibinfo {author} {\bibfnamefont {A.}~\bibnamefont
  {Delteil}}, \bibinfo {author} {\bibfnamefont {Z.}~\bibnamefont {Sun}},
  \bibinfo {author} {\bibfnamefont {S.}~\bibnamefont {F\"alt}}, \ and\ \bibinfo
  {author} {\bibfnamefont {A.}~\bibnamefont {Imamo\ifmmode~\breve{g}\else
  \u{g}\fi{}lu}},\ }\href {\doibase 10.1103/PhysRevLett.118.177401} {\bibfield
  {journal} {\bibinfo  {journal} {Phys. Rev. Lett.}\ }\textbf {\bibinfo
  {volume} {118}},\ \bibinfo {pages} {177401} (\bibinfo {year}
  {2017})}\BibitemShut {NoStop}%
\bibitem [{\citenamefont {Childress}\ \emph {et~al.}(2005)\citenamefont
  {Childress}, \citenamefont {Taylor}, \citenamefont {S\o{}rensen},\ and\
  \citenamefont {Lukin}}]{Childress2005A}%
  \BibitemOpen
  \bibfield  {author} {\bibinfo {author} {\bibfnamefont {L.}~\bibnamefont
  {Childress}}, \bibinfo {author} {\bibfnamefont {J.~M.}\ \bibnamefont
  {Taylor}}, \bibinfo {author} {\bibfnamefont {A.~S.}\ \bibnamefont
  {S\o{}rensen}}, \ and\ \bibinfo {author} {\bibfnamefont {M.~D.}\ \bibnamefont
  {Lukin}},\ }\href {\doibase 10.1103/PhysRevA.72.052330} {\bibfield  {journal}
  {\bibinfo  {journal} {Phys. Rev. A}\ }\textbf {\bibinfo {volume} {72}},\
  \bibinfo {pages} {052330} (\bibinfo {year} {2005})}\BibitemShut {NoStop}%
\bibitem [{\citenamefont {Maurer}\ \emph {et~al.}(2012)\citenamefont {Maurer},
  \citenamefont {Kucsko}, \citenamefont {Latta}, \citenamefont {Jiang},
  \citenamefont {Yao}, \citenamefont {Bennett}, \citenamefont {Pastawski},
  \citenamefont {Hunger}, \citenamefont {Chisholm}, \citenamefont {Markham},
  \citenamefont {Twitchen}, \citenamefont {Cirac},\ and\ \citenamefont
  {Lukin}}]{Maurer2012}%
  \BibitemOpen
  \bibfield  {author} {\bibinfo {author} {\bibfnamefont {P.~C.}\ \bibnamefont
  {Maurer}}, \bibinfo {author} {\bibfnamefont {G.}~\bibnamefont {Kucsko}},
  \bibinfo {author} {\bibfnamefont {C.}~\bibnamefont {Latta}}, \bibinfo
  {author} {\bibfnamefont {L.}~\bibnamefont {Jiang}}, \bibinfo {author}
  {\bibfnamefont {N.~Y.}\ \bibnamefont {Yao}}, \bibinfo {author} {\bibfnamefont
  {S.~D.}\ \bibnamefont {Bennett}}, \bibinfo {author} {\bibfnamefont
  {F.}~\bibnamefont {Pastawski}}, \bibinfo {author} {\bibfnamefont
  {D.}~\bibnamefont {Hunger}}, \bibinfo {author} {\bibfnamefont
  {N.}~\bibnamefont {Chisholm}}, \bibinfo {author} {\bibfnamefont
  {M.}~\bibnamefont {Markham}}, \bibinfo {author} {\bibfnamefont {D.~J.}\
  \bibnamefont {Twitchen}}, \bibinfo {author} {\bibfnamefont {J.~I.}\
  \bibnamefont {Cirac}}, \ and\ \bibinfo {author} {\bibfnamefont {M.~D.}\
  \bibnamefont {Lukin}},\ }\href {\doibase 10.1126/science.1220513} {\bibfield
  {journal} {\bibinfo  {journal} {Science}\ }\textbf {\bibinfo {volume}
  {336}},\ \bibinfo {pages} {1283} (\bibinfo {year} {2012})}\BibitemShut
  {NoStop}%
\bibitem [{\citenamefont {Wolters}\ \emph {et~al.}(2017)\citenamefont
  {Wolters}, \citenamefont {Buser}, \citenamefont {Horsley}, \citenamefont
  {B\'eguin}, \citenamefont {J\"ockel}, \citenamefont {Jahn}, \citenamefont
  {Warburton},\ and\ \citenamefont {Treutlein}}]{Wolters2017}%
  \BibitemOpen
  \bibfield  {author} {\bibinfo {author} {\bibfnamefont {J.}~\bibnamefont
  {Wolters}}, \bibinfo {author} {\bibfnamefont {G.}~\bibnamefont {Buser}},
  \bibinfo {author} {\bibfnamefont {A.}~\bibnamefont {Horsley}}, \bibinfo
  {author} {\bibfnamefont {L.}~\bibnamefont {B\'eguin}}, \bibinfo {author}
  {\bibfnamefont {A.}~\bibnamefont {J\"ockel}}, \bibinfo {author}
  {\bibfnamefont {J.-P.}\ \bibnamefont {Jahn}}, \bibinfo {author}
  {\bibfnamefont {R.~J.}\ \bibnamefont {Warburton}}, \ and\ \bibinfo {author}
  {\bibfnamefont {P.}~\bibnamefont {Treutlein}},\ }\href {\doibase
  10.1103/PhysRevLett.119.060502} {\bibfield  {journal} {\bibinfo  {journal}
  {Phys. Rev. Lett.}\ }\textbf {\bibinfo {volume} {119}},\ \bibinfo {pages}
  {060502} (\bibinfo {year} {2017})}\BibitemShut {NoStop}%
\bibitem [{\citenamefont {Collins}\ \emph {et~al.}(2007)\citenamefont
  {Collins}, \citenamefont {Jenkins}, \citenamefont {Kuzmich},\ and\
  \citenamefont {Kennedy}}]{Collins2007}%
  \BibitemOpen
  \bibfield  {author} {\bibinfo {author} {\bibfnamefont {O.~A.}\ \bibnamefont
  {Collins}}, \bibinfo {author} {\bibfnamefont {S.~D.}\ \bibnamefont
  {Jenkins}}, \bibinfo {author} {\bibfnamefont {A.}~\bibnamefont {Kuzmich}}, \
  and\ \bibinfo {author} {\bibfnamefont {T.~A.~B.}\ \bibnamefont {Kennedy}},\
  }\href {\doibase 10.1103/PhysRevLett.98.060502} {\bibfield  {journal}
  {\bibinfo  {journal} {Phys. Rev. Lett.}\ }\textbf {\bibinfo {volume} {98}},\
  \bibinfo {pages} {060502} (\bibinfo {year} {2007})}\BibitemShut {NoStop}%
\bibitem [{\citenamefont {Muralidharan}\ \emph {et~al.}(2014)\citenamefont
  {Muralidharan}, \citenamefont {Kim}, \citenamefont {L\"utkenhaus},
  \citenamefont {Lukin},\ and\ \citenamefont {Jiang}}]{Muralidharan2014}%
  \BibitemOpen
  \bibfield  {author} {\bibinfo {author} {\bibfnamefont {S.}~\bibnamefont
  {Muralidharan}}, \bibinfo {author} {\bibfnamefont {J.}~\bibnamefont {Kim}},
  \bibinfo {author} {\bibfnamefont {N.}~\bibnamefont {L\"utkenhaus}}, \bibinfo
  {author} {\bibfnamefont {M.~D.}\ \bibnamefont {Lukin}}, \ and\ \bibinfo
  {author} {\bibfnamefont {L.}~\bibnamefont {Jiang}},\ }\href {\doibase
  10.1103/PhysRevLett.112.250501} {\bibfield  {journal} {\bibinfo  {journal}
  {Phys. Rev. Lett.}\ }\textbf {\bibinfo {volume} {112}},\ \bibinfo {pages}
  {250501} (\bibinfo {year} {2014})}\BibitemShut {NoStop}%
\bibitem [{\citenamefont {Pearle}(1970)}]{Pearle1970}%
  \BibitemOpen
  \bibfield  {author} {\bibinfo {author} {\bibfnamefont {P.~M.}\ \bibnamefont
  {Pearle}},\ }\href {\doibase 10.1103/PhysRevD.2.1418} {\bibfield  {journal}
  {\bibinfo  {journal} {Phys. Rev. D}\ }\textbf {\bibinfo {volume} {2}},\
  \bibinfo {pages} {1418} (\bibinfo {year} {1970})}\BibitemShut {NoStop}%
\bibitem [{\citenamefont {Larsson}(2014)}]{Larsson2014}%
  \BibitemOpen
  \bibfield  {author} {\bibinfo {author} {\bibfnamefont {J.-{\AA}.}\
  \bibnamefont {Larsson}},\ }\href
  {http://stacks.iop.org/1751-8121/47/i=42/a=424003} {\bibfield  {journal}
  {\bibinfo  {journal} {Journal of Physics A: Mathematical and Theoretical}\
  }\textbf {\bibinfo {volume} {47}},\ \bibinfo {pages} {424003} (\bibinfo
  {year} {2014})}\BibitemShut {NoStop}%
\bibitem [{\citenamefont {Ewert}\ \emph {et~al.}(2016)\citenamefont {Ewert},
  \citenamefont {Bergmann},\ and\ \citenamefont {van Loock}}]{Ewert2016}%
  \BibitemOpen
  \bibfield  {author} {\bibinfo {author} {\bibfnamefont {F.}~\bibnamefont
  {Ewert}}, \bibinfo {author} {\bibfnamefont {M.}~\bibnamefont {Bergmann}}, \
  and\ \bibinfo {author} {\bibfnamefont {P.}~\bibnamefont {van Loock}},\ }\href
  {\doibase 10.1103/PhysRevLett.117.210501} {\bibfield  {journal} {\bibinfo
  {journal} {Phys. Rev. Lett.}\ }\textbf {\bibinfo {volume} {117}},\ \bibinfo
  {pages} {210501} (\bibinfo {year} {2016})}\BibitemShut {NoStop}%
\bibitem [{\citenamefont {Ewert}\ and\ \citenamefont {van
  Loock}(2017)}]{Ewert2017}%
  \BibitemOpen
  \bibfield  {author} {\bibinfo {author} {\bibfnamefont {F.}~\bibnamefont
  {Ewert}}\ and\ \bibinfo {author} {\bibfnamefont {P.}~\bibnamefont {van
  Loock}},\ }\href {\doibase 10.1103/PhysRevA.95.012327} {\bibfield  {journal}
  {\bibinfo  {journal} {Phys. Rev. A}\ }\textbf {\bibinfo {volume} {95}},\
  \bibinfo {pages} {012327} (\bibinfo {year} {2017})}\BibitemShut {NoStop}%
\bibitem [{\citenamefont {Lee}\ \emph {et~al.}(2018)\citenamefont {Lee},
  \citenamefont {Ralph},\ and\ \citenamefont {Jeong}}]{Lee2018}%
  \BibitemOpen
  \bibfield  {author} {\bibinfo {author} {\bibfnamefont {S.-W.}\ \bibnamefont
  {Lee}}, \bibinfo {author} {\bibfnamefont {T.~C.}\ \bibnamefont {Ralph}}, \
  and\ \bibinfo {author} {\bibfnamefont {H.}~\bibnamefont {Jeong}},\
  }\href@noop {} {\bibfield  {journal} {\bibinfo  {journal} {arXiv:1804.09342}\
  } (\bibinfo {year} {2018})}\BibitemShut {NoStop}%
\bibitem [{\citenamefont {Glaudell}\ \emph {et~al.}(2016)\citenamefont
  {Glaudell}, \citenamefont {Waks},\ and\ \citenamefont
  {Taylor}}]{Glaudell2016}%
  \BibitemOpen
  \bibfield  {author} {\bibinfo {author} {\bibfnamefont {A.~N.}\ \bibnamefont
  {Glaudell}}, \bibinfo {author} {\bibfnamefont {E.}~\bibnamefont {Waks}}, \
  and\ \bibinfo {author} {\bibfnamefont {J.~M.}\ \bibnamefont {Taylor}},\
  }\href {http://stacks.iop.org/1367-2630/18/i=9/a=093008} {\bibfield
  {journal} {\bibinfo  {journal} {New Journal of Physics}\ }\textbf {\bibinfo
  {volume} {18}},\ \bibinfo {pages} {093008} (\bibinfo {year}
  {2016})}\BibitemShut {NoStop}%
\bibitem [{\citenamefont {Muralidharan}\ \emph {et~al.}(2017)\citenamefont
  {Muralidharan}, \citenamefont {Zou}, \citenamefont {Li}, \citenamefont
  {Wen},\ and\ \citenamefont {Jiang}}]{Muralidharan2017}%
  \BibitemOpen
  \bibfield  {author} {\bibinfo {author} {\bibfnamefont {S.}~\bibnamefont
  {Muralidharan}}, \bibinfo {author} {\bibfnamefont {C.-L.}\ \bibnamefont
  {Zou}}, \bibinfo {author} {\bibfnamefont {L.}~\bibnamefont {Li}}, \bibinfo
  {author} {\bibfnamefont {J.}~\bibnamefont {Wen}}, \ and\ \bibinfo {author}
  {\bibfnamefont {L.}~\bibnamefont {Jiang}},\ }\href
  {http://stacks.iop.org/1367-2630/19/i=1/a=013026} {\bibfield  {journal}
  {\bibinfo  {journal} {New Journal of Physics}\ }\textbf {\bibinfo {volume}
  {19}},\ \bibinfo {pages} {013026} (\bibinfo {year} {2017})}\BibitemShut
  {NoStop}%
\bibitem [{\citenamefont {Muralidharan}\ \emph {et~al.}(2018)\citenamefont
  {Muralidharan}, \citenamefont {Zou}, \citenamefont {Li},\ and\ \citenamefont
  {Jiang}}]{Muralidharan2018}%
  \BibitemOpen
  \bibfield  {author} {\bibinfo {author} {\bibfnamefont {S.}~\bibnamefont
  {Muralidharan}}, \bibinfo {author} {\bibfnamefont {C.-L.}\ \bibnamefont
  {Zou}}, \bibinfo {author} {\bibfnamefont {L.}~\bibnamefont {Li}}, \ and\
  \bibinfo {author} {\bibfnamefont {L.}~\bibnamefont {Jiang}},\ }\href
  {\doibase 10.1103/PhysRevA.97.052316} {\bibfield  {journal} {\bibinfo
  {journal} {Phys. Rev. A}\ }\textbf {\bibinfo {volume} {97}},\ \bibinfo
  {pages} {052316} (\bibinfo {year} {2018})}\BibitemShut {NoStop}%
\end{thebibliography}
%
	\clearpage

\end{document}